\documentclass[]{aastex631}

\shorttitle{Rot. Vel. in NGC 2264 T Tauri Stars}
\shortauthors{Gray et al.}

\graphicspath{{./}{figures/}}

\begin{document}

\title{Rotational Velocities and Radii Estimates of Low-Mass Pre-Main Sequence Stars in NGC 2264}

\author[0000-0001-6389-5639]{Laurin M. Gray}
\affiliation{Department of Astronomy, Indiana University, 727 East Third Street, Bloomington, IN 47405, USA}
\email{grayla@iu.edu}

\author[0000-0001-8283-4591]{Katherine L. Rhode}
\affiliation{Department of Astronomy, Indiana University, 727 East Third Street, Bloomington, IN 47405, USA}

\author[0000-0001-5924-3531]{Catrina M. Hamilton-Drager}
\affiliation{Department of Physics and Astronomy, Dickinson College, 28 N. College Street, Carlisle, PA 17013, USA}

\author{Tiffany Picard}
\affiliation{Astronomy Department, Mount Holyoke College, 50 College Street, South Hadley, MA 01075, USA}

\author[0000-0001-6381-515X]{Luisa M. Rebull}
\affiliation{Infrared Science Archive (IRSA), IPAC, California Institute of Technology, 1200 E. California Blvd., Pasadena, CA 91125, USA}

\begin{abstract}

Investigating the angular momentum evolution of pre-main sequence (PMS) stars provides important insight into the interactions between Sun-like stars and their protoplanetary disks, and the timescales that govern disk dissipation and planet formation. We present projected rotational velocities ($v$~sin~$i$ values) of 254 T Tauri stars (TTSs) in the $\sim$3 Myr-old open cluster NGC 2264, measured using high-dispersion spectra from the WIYN 3.5m telescope's Hydra instrument. We combine these with literature values of temperature, rotation period, luminosity, disk classification, and binarity. We find some evidence that Weak-lined TTSs may rotate faster than their Classical TTS counterparts and that stars in binary systems may rotate faster than single stars. We also combine our $v$~sin~$i$ measurements with rotation period to estimate the projected stellar radii of our sample stars, and then use a maximum likelihood modeling technique to compare our radii estimates to predicted values from stellar evolution models. We find that starspot-free models tend to underestimate the radii of the PMS stars at the age of the cluster, while models that incorporate starspots are more successful. We also observe a mass dependence in the degree of radius inflation, which may be a result of differences in the birthline location on the HR diagram. Our study of NGC 2264 serves as a pilot study for analysis methods to be applied to four other clusters ranging in age from 1 to 14 Myr, which is the timescale over which protoplanetary disks dissipate and planetary systems begin to form.

\end{abstract}


\section{Introduction} \label{sec:intro}

An important part of developing our picture of the Solar System's formation is understanding the evolution of young Sun-like stars and their surrounding protoplanetary disks, particularly during the pre-main sequence (PMS) phase when the star and disk are still in the process of forming and are closely linked to each other. A key parameter during the pre-main sequence stage is the angular momentum of the star. Like other fundamental stellar properties such as mass and metallicity, a star's rotation rate is an important characteristic that affects its internal structure and evolution during the pre-main sequence phase and throughout its entire lifetime. Rotation affects chemical mixing, energy transport within the star, and stellar winds and mass loss (e.g., \citealt{Pinsonneault1989, Pinsonneault1990, Barnes2003, Ekstrom2012, Bouvier2014}; and many others).  

T Tauri stars are pre-main sequence stars with masses $\lesssim$ 2$-$3 $M_{\odot}$, and very young ages (typically $\lesssim$ $10^7$ years) (e.g., \citealt{Bertout1989, Herbst1994}).  Over the past few decades, observational and theoretical studies have enabled us to develop a general picture of how angular momentum evolution proceeds in T Tauri stars and how the stars interact and co-evolve with their protoplanetary disks.  The T Tauri stage is one of rapid evolution; bipolar outflows and jets expel stellar material, while magnetic fields produce cooler, darker starspots on the stellar surface and channel gas to the surface from the circumstellar disk along the magnetic field lines to create brighter hot spots (e.g., \citealt{Mundt1983, Lada1985, Konigl1991, Shu1994a, Edwards1994, Bachiller1996, Feigelson1999, Cody2014}). One useful approach is to study T Tauri stars in nearby star-forming regions in order to investigate and quantify the stars' properties in detail.  

Measuring rotation can provide us with geometric estimates of radii for stars, by combining the projected rotation velocity with the rotation period (e.g., \citealt{Rhode2001b}).  These radii are not dependent on uncertainties in temperature or luminosity, and so represent an additional tool for measuring radii.  For an individual star, due to the dependence of the projected rotation velocity measurement on the inclination $i$ of the star's spin-axis with respect to the observer (which is unknown), this gives a lower limit for the radius.  With measurements of many stars, the estimates can be compared to predictions from stellar evolutionary models through statistical modeling methods \citep{Lanzafame2017, Jackson2018}.  Measurements of radius and mass from PMS and zero-age main sequence (ZAMS) eclipsing binary stars indicate that models without starspots tend to underestimate the radii  (e.g., \citealt{Lopez-Morales2007, Torres2010, Kraus2015, David2019, Smith2021}). The high magnetic activity of PMS stars can pose an additional challenge for these models, as starspots can affect the effective temperature and luminosity of the star \citep{Somers2015a, Somers2020}. 

Rotation studies of many young star clusters have revealed that the rotation speeds of T Tauri stars are often slower than predicted, indicating significant angular momentum loss at a very early stage \citep{Vogel1981, Hartmann1986, Clarke2000}.  Some populations also show a bimodal distribution of rotation periods, indicating that not all stars are affected by the same processes (e.g., \citealt{Nordhagen2006, Cieza2006, Venuti2017}).  T Tauri stars are commonly separated into two classes according to their accretion properties and, by extension, their circumstellar disk properties. Classical T Tauri Stars (CTTSs) show strong emission lines in their spectra and signatures of mass accretion from a circumstellar disk to the star, such as hot spots on their surfaces that can cause flaring and excess emission in optical, UV, X-ray, and/or IR wavebands (e.g., \citealt{Bertout1989}). By contrast, Weak-lined T Tauri Stars (WTTSs) often show periodic variability, likely due to large surface cool spots, and lack obvious accretion disk signatures like NIR excess and Ca II emission lines (e.g., \citealt{Walter1987, Herbst1994, Hillenbrand1998b}). \cite{Briceno2019} have also introduced a transitory classification of CWTTS, representing the scenario where the disk may be present but weak, possibly in the process of becoming truncated.  Some have linked the bimodal rotation period distribution to these two classes of stars, with the slower rotation of some stars possibly being connected to the presence of an accreting disk (e.g., \citealt{Edwards1993, Herbst2001, Herbst2002, Cieza2007b, Venuti2017, Rebull2018, Serna2021}). However, further investigation is required to fully understand the links between disks and rotation, as well as to quantify the time scales and physical conditions that govern disk dissipation. One proposed theory is that a magnetic interaction between the star and its surrounding disk regulates the star's rotation rate, removing angular momentum and slowing the star down (e.g., \citealt{Konigl1991, Shu1994a, Ostriker1995, Matt2005b, Bouvier2014}). This mechanism is often referred to as ``disk-locking", in which kilogauss-strength magnetic fields, threaded through the circumstellar disk, connect the disk to the star's surface and channel accreting material onto the star. The star is locked to the disk by this magnetic interaction and forced to have the same rotation rate as the disk at the locking radius. Only when the disk disperses does the star become unlocked and spin up, so a star that is disk-locked (or only recently unlocked) will have a slow rotation period, whereas a fast-rotating star may either have never been disk-locked or has had time to spin up since its disk dispersed (e.g., \citealt{Bouvier1993, Edwards1993, Rebull2004, Bouvier2014}). The expected time scale over which the disk disperses and the star spins up is estimated to be anywhere from $\sim$1 to 5 Myr (e.g., \citealt{Herbst2000, Nordhagen2006, Cieza2007a, Hartmann2016, Bastian2020, Rebull2020}). These time scales are uncertain, however, and some recent work suggests that disk lifetimes could be as long as 5$-$10 Myr \citep{Fedele2010, Gallet2015, Pfalzner2022, Pfalzner2024}.

To investigate rotation in T Tauri stars during this early and fast-changing stage in their evolution, we have initiated a study measuring $v$~sin~$i$ for a large sample of low-mass PMS stars in open clusters with ages that span 1$-$14 Myr.  This age range will allow us to explore evolution during and shortly after the era of the circumstellar disk.

NGC 2264 is a young open cluster that is well-researched and makes an ideal target for a pilot study to develop our analysis methods.  \cite{Sung1997} estimated a distance of $\sim$760 $\pm$ 49 pc from fitting the ZAMS, which has been consistent with later estimates using a variety of methods (e.g. isochrone fitting: \citealt{Turner2012}; parallax-based: \citealt{Kamezaki2013}; PMS eclipsing binaries: \citealt{Gillen2020}).  NGC 2264 has minimal foreground extinction and a well-defined population, including a large sample of confirmed PMS members \citep{Venuti2018}.  An age of up to 7 Myr has been suggested, though estimates usually range from 3$-$5 Myr \citep{Sung1997, Park2000, Venuti2018, Dias2021}.  \cite{Flaccomio2000} and \cite{Ramirez2004} have proposed that an age of 1$-$3 Myr may be more appropriate, based on measurements of lower mass stars.  \cite{Venuti2018, Venuti2019} and \cite{Lim2016} have also explored the possibility of an age spread; \cite{Venuti2019} estimated an age range of $\sim$0.5$-$5 Myr for the cluster populations while \cite{Lim2016} proposed a narrower spread of 3$-$4 Myr with a total formation timescale $<$ 5 Myr.  For this work, our starting assumption is an average age of 3 Myr.

We have collected spectra for $\sim$300 stars in NGC 2264 with the WIYN 3.5m telescope at Kitt Peak National Observatory (KPNO), in order to measure their $v$~sin~$i$.  In Section \ref{sec:sample-and-obs}, we discuss our target selection and observations and in Section \ref{sec:analysis}, we cover the measurement and analysis methods.  Section \ref{sec:results} contains our results and the related discussion.  In Section \ref{sec:conclusion}, we summarize our findings.

\section{Observations and Data Reduction} \label{sec:sample-and-obs}

\subsection{Sample Selection} \label{sec:sample}

We relied heavily on previously published auxiliary data such as sky positions, photometric magnitude, color, and effective temperature to select an appropriate target sample of low-mass PMS stars to observe with WIYN Hydra.  We began by building a comprehensive database of all PMS stars in NGC 2264 from which to select our observational sample.  The basis of our catalog was 164 objects which had previously been observed with Hydra at the WIYN 3.5m telescope, but the spectra had never been published (see Section \ref{sec:obs}).  These objects were originally selected as low-mass PMS NGC 2264 members by their magnitude, color, and proper motion from \cite{Jones1988}. We then added objects identified as PMS NGC 2264 members from \cite{Lamm2004, Lamm2005}, \cite{Rebull2002a}, \cite{Venuti2014, Venuti2017, Venuti2018, Venuti2019}, \cite{Baxter2009}, and \cite{Affer2013}.  These catalogs contained photometry, spectral type, effective temperature, rotation period, rotational velocity, and/or disk classifications.  For objects with photometry from \cite{Venuti2014, Venuti2017, Venuti2018, Venuti2019}, we converted $gri-$band photometry to $I_{\rm C}$ and $V$ band using the conversions of \cite{Jester2005}.  Each new catalog was cross-matched against the previously added objects using a 2$\arcsec$ search radius.  We manually checked for sources that had been mistakenly duplicated and confirmed objects that had been cross-matched, relying on photometry and previous cross-matching done in \cite{Lamm2004} and on Simbad \citep{SIMBAD2000}.  We also excluded objects that were marked as being part of multi-component systems in \cite{Venuti2017}, and any objects to which they were cross-matched.  Finally, we added additional rotation periods \citep{Makidon2004}, rotational velocities \citep{Baxter2009, Jackson2016}, and radial velocities \citep{Furesz2006, Jackson2016} for objects that were already in the database.

Creating a catalog by combining several different sources means that the sky coordinates do not share a common astrometric solution, but multi-object spectroscopy requires precise astrometry for fiber placement.  The Gaia EDR3 sky survey provides sky positions for billions of stars with a brightness limit well below those of our targets, so we were confident that it could provide a common coordinate source for our objects \citep{GaiaEDR3-2021a}.  We performed a cross-match within 2$\arcsec$ to the Gaia EDR3 archive and replaced all object sky coordinates with the ones from Gaia EDR3.  Objects that did not have a match within 2$\arcsec$ were removed from the database.  When multiple Gaia objects were identified as a potential match, the objects were checked in Simbad \citep{SIMBAD2000}. If Simbad classified them as part of a multi-component system, the corresponding component was identified in the database and noted; if not, the coordinates of the closest Gaia object were chosen.  One complication is that objects that were identified as separate sources in Gaia EDR3 were often reported as singular objects by the original catalogs, and it is impossible to know whether the reported measurements pertain to the same component (and which one it is).  For this reason, we attempted to identify and exclude any such objects from the potential target list.  

To build our list of potential targets, we selected all objects with the following conditions in at least one of the source catalogs: (1) $I_{\rm C}$ magnitude from 10 to 16, (2) $V - I_{\rm C}$ between 1 and 4, and (3) a spectral type or effective temperature indicating a low-mass PMS star ($T_{\rm eff} \leq$ 4760 K).  According to the pre-main sequence evolutionary models of \cite{Baraffe2015}, 4760 K is roughly the temperature of a 1.4 $M_{\odot}$ star at 5 Myr.  As the accepted age estimates for NGC 2264 generally range from 3$-$5 Myr, this therefore represents the expected upper temperature limit of low-mass stars in the cluster.  The Hydra fiber assignment program, whydra, allows for weighted prioritization of targets when creating a pointing file for multi-object spectroscopy.  We prioritized objects with a reported period followed by those with reported luminosity; for stars with both, we are able to estimate the radius and equatorial velocity of the star.  Fainter stars required a longer integration time, so we split the potential target list into a ``bright" sample and a ``faint" sample at $I_{\rm C}$ = 14.75 mag.  

\subsection{Adopted Values from the Literature} \label{sec:adoption}
Great care must be taken when combining catalogs and selecting reported literature values.  Some objects may have multiple, differing measurements reported for a single property, such as rotation period.  Some papers may not clearly state the source for reported data on individual objects, leading to repeated results when catalogs are combined.  This means that simply taking a mode or average of all reported values for a given property is not appropriate.  Instead, we adopted reported auxiliary data as follows.  

For spectral types and temperatures, we used those reported by \cite{Venuti2014}, followed by types from \cite{Rebull2002a}, photometrically estimated types in \cite{Venuti2014}, and effective temperatures from \cite{Venuti2018}.  When a spectral type was the preferred selection for a star, we converted it to a $T_{\rm eff}$ following the prescription of \cite{Pecaut2013}, which takes into account that PMS stars may have cooler surface temperatures than a main sequence star of corresponding spectral type. In the case where a listed spectral type was between types (e.g., M2-3), we used the spectral type in the middle of the range (e.g., M2.5) and calculated a temperature between the two nearest types.  When a type was given as ``$>$M4", we considered it M5 as this was the latest type with a temperature conversion for PMS stars in \cite{Pecaut2013}.  For periods, we selected the space-based period measurements from \cite{Venuti2017} and \cite{Affer2013}, followed by periods from \cite{Lamm2005} and \cite{Makidon2004}.  We selected luminosities reported in \cite{Venuti2018} over those from \cite{Jackson2016} when possible.  Radius was calculated using the adopted $T_{\rm eff}$ and $L_{\rm bol}$, but if the $L_{\rm bol}$ was not available, we adopted the radius from \cite{Rebull2002a} if one was listed.

Multiple papers report results on whether stars are CTTSs or WTTSs, or if they are actively accreting, though they use different diagnostic methods and terminology.  Some papers \citep{Furesz2006, Affer2013} use active accretion as a proxy for disk presence, while \cite{Venuti2018} distinguishes between disk-bearing and actively accreting.  Furthermore, T Tauri stars may go through periods of low or inactive accretion, and so may be classified as non-accreting at one point in time and accreting at another \citep{Venuti2018}.  \cite{Venuti2014, Venuti2018} and \cite{Jackson2016} used IR excess to indicate the presence of a dusty disk, while \cite{Furesz2006}, \cite{Affer2013}, and \cite{Venuti2018} used UV excess, the $H\alpha$ equivalent width, and/or a threshold of 270 $\mathrm{km}\,\mathrm{s}^{-1}$ for the W10\%($H\alpha$) parameter to evaluate whether an object was actively accreting.  In \cite{Venuti2018}, it was highly unusual for an object to be accreting without a corresponding disk-bearing identification, so we consider active accretion to be an indicator of a disk-bearing star.    

In light of the different test types, nomenclatures, and observation time periods, we devised the following schema for identifying an object based on the literature: \cite{Affer2013} and \cite{Venuti2018} used different thresholds for the $H\alpha$ equivalent width to classify their objects, so to achieve consistency in our classification, we reclassified their data using the system defined by \cite{Briceno2019}.  \cite{Briceno2019} uses spectral type-dependent thresholds to classify stars as CTTS, WTTS, or a new, intermediate class called CWTTS.  This class may represent a transition state between the CTTS and WTTS stages, where the disk is very weak or truncated \citep{Briceno2019}.  If the $H\alpha$ equivalent width in \cite{Affer2013} or \cite{Venuti2018} placed an object in CWTTS, it was considered a CWTTS.  Otherwise, if the data in \cite{Furesz2006}, \cite{Affer2013}, or \cite{Venuti2014, Venuti2018} indicated that an object had a disk or was accreting, it was considered a CTTS.  For the remaining objects, if any of those tests indicated an object was not a CTTS, it was marked WTTS.  Finally, in the absence of any other information, an IR excess in \cite{Jackson2016} was used to classify a star as CTTS.  \cite{Jackson2016} was considered separately because they frequently indicated an IR excess in disagreement with \cite{Venuti2014, Venuti2018}, despite using data from the same time period. \cite{Venuti2014, Venuti2018} often validated their non-disk results with an accretion test that indicated a lack of active accretion, so we considered them to be more reliable than other papers that did not have a secondary validation.  

\subsection{Observations} \label{sec:obs}

All observations were completed using the Bench spectrograph and Hydra multi-fiber positioner on the WIYN 3.5m telescope at Kitt Peak National Observatory (KPNO).  Hydra allows for simultaneous spectroscopy of up to $\sim$100 objects over a 1$^\circ$ area of sky.  We used the red fibers on Hydra, paired with the Bench Camera and 316@63.4 echelle grating.  This allowed us to balance the desire for high resolution (R $\sim$ 21,500) against our ability to observe fainter objects (I $\sim$ 16) with integration times between three and four hours. The spectra were centered on 6400 {\AA} with a dispersion of 0.145 {\AA} per pixel, leading to a spectral range of 6240 to 6450 {\AA}.  A Thorium-Argon comparison lamp spectrum was taken at minimum before and after the set of exposures for each configuration, and occasionally partway through.  

The initial set of observations was conducted in 1997 December, 1998 December, and 2002 January \citep{Hamilton2007}.  Observations were done in four different configurations, with four 40-minute exposures for each configuration.  Each configuration contained 50$-$60 stars and at least 15$-$20 fibers placed on sky positions.  These observations comprise the spectra for the 164 objects we used as the basis of our database.

We conducted further observations in 2022 December, 2023 February, and 2023 December.  
Observations were taken in six different configurations, with six 30-minute exposures for ``bright" configurations and seven 30-minute exposures for ``faint" configurations.  Each configuration contained 50$-$60 stars and at least nine sky fibers.  

In total, spectra were obtained for 312 objects.  We observed 89 stars more than once, which allowed us to compare radial velocities at multiple epochs and to average measurements.  We also observed seven slow-rotating stars with known spectral types (G6V$-$M2V) and the Sun to serve as narrow-lined templates for cross-correlation, and ten bright stars with previously measured rotational velocities from \cite{Mermilliod2009} (ranging from 9.9 to 35.1 $\mathrm{km}\,\mathrm{s}^{-1}$) to help characterize our lower limits.  

\subsection{Data Reduction}

The spectroscopic data were reduced using IRAF \citep{Tody1986, Tody1993} tasks such as ZEROCOMBINE, DARKCOMBINE, and FLATCOMBINE to create combined bias, dark, and dome flat images, and CCDPROC was used for bias-subtraction, dark-subtraction, and trimming on non-zero and non-dark frames.  The IRAF task DOHYDRA was used to extract apertures, apply the combined dome flat, correct for scattered light, calibrate the wavelength dispersion function, and subtract the sky for the object spectra.  Observations of the ThAr spectrum were used for wavelength calibration, with exposures taken within $\sim$ 3 hours of any given science observation.  For the sky subtraction, sky fibers were examined for unusually high signal (indicating that they had flux from an object) and removed.  The remaining sky spectra were averaged to make a combined sky spectrum, with cosmic rays removed by a sigma-clipping algorithm, which was then subtracted from the object spectra.  Due to the field size allowed by Hydra ($\sim$1$^{\circ}$) and the random distribution of sky positions, sky fibers may collect flux from both the sky and NGC 2264.  As noted in \cite{Rhode2001b}, this variable background should not affect the measurement of the rotational velocity.  

After the object spectra were processed with DOHYDRA, the individual exposures of each configuration were scaled, weighted, and averaged into a single image using the IRAF task SCOMBINE.  This process removes cosmic rays and improves the signal-to-noise of the spectra.  While some stars were observed multiple times between various configurations, we treated these as independent observations and did not combine spectra across configurations.  We clipped the spectra so that they covered the region from 6275 {\AA} to 6525 {\AA}, removing regions with poorer focus on the ends of the spectra, and continuum-normalized them by fitting a second order spline3 function with the CONTINUUM task.

Reductions for observations of the narrow-lined template stars were conducted the same way, with the exception of continuum-normalization, which was done as part of the cross-correlation task FXCOR.

\section{Data Analysis} \label{sec:analysis}

\subsection{Measuring Radial and Rotational Velocities}

Radial and rotational velocities of the stars were measured with the IRAF task FXCOR, which performs a Fourier cross-correlation between an object spectrum and a template spectrum.  When a stellar spectrum is cross-correlated against a narrow-lined spectrum with a known radial velocity, the location of the cross-correlation function (CCF) peak gives the radial velocity (RV) of the object, while the width of the peak is related to the amount that the absorption lines have been rotationally broadened.  By measuring the full width at half maximum (FWHM) of the CCF peak, we can measure the rotational velocity of the star.  The measured velocity is subject to the angle between the observer and the star's rotational axis, so that this is actually a measurement of the projected rotational velocity, or $v$~sin~$i$.  

To produce the clearest cross-correlations, each science star should be cross-correlated against a narrow-lined template for a star with a similar temperature.  To achieve this, we obtained spectra for seven different slow-rotating stars with spectral types from G6V to M2V, plus the Sun. Information about the narrow-lined template stars is reported in Table \ref{tab:templates}.  We converted these spectral types to effective temperatures using the reported values from \cite{Pecaut2013} for dwarf stars.  We then matched each object in our sample to the narrow-lined template that was closest in temperature.  

\begin{deluxetable}{cccccc}
\tablecaption{Bright, slow-rotating stars observed for use as narrow-lined spectral templates for cross-correlation}
\tablehead{\colhead{Star} & \colhead{RA} & \colhead{Dec} & \colhead{SpT} & \colhead{log($T_{\rm eff}$)} & \colhead{V} \\ 
\colhead{} & \colhead{(deg)} & \colhead{(deg)} & \colhead{} & \colhead{} & \colhead{(mag)} } 

\startdata
Sun & \nodata & \nodata & G2V & 3.761 & -26 \\
Gl 59.1 & 23.63392 & 68.94865 & G6V & 3.747 & 6.5 \\
Gl 75 & 26.94268 & 63.85141 & K0V & 3.723 & 5.6 \\
Gl 144 & 53.22829 & -9.45817 & K2V & 3.702 & 3.7 \\
Gl 53.1A & 16.90831 & 22.95280 & K4V & 3.665 & 8.4 \\
Gl 114 & 42.65531 & 15.70816 & K6V & 3.623 & 8.9 \\
Gl 116 & 43.03510 & 34.38495 & M0V & 3.585 & 9.6 \\
Gl 411 & 165.83096 & 35.94865 & M2V & 3.550 & 7.5 \\
\enddata

\label{tab:templates}
\tablecomments{Right ascension and declination are given in degrees, using coordinates from Gaia DR3 \citep{GaiaEDR3-2021a}.  The sky coordinates for the Sun are variable.  Temperatures are derived from spectral types using the dwarf SpT-$T_{\rm eff}$ relations in \cite{Pecaut2013}.}
\end{deluxetable}

In order to measure the $v$~sin~$i$ of a star, we must understand the relationship between the velocity and the FWHM of the cross-correlation peak. To accomplish this, we artificially ``spun up" each narrow-lined template by convolving it with a theoretical rotation profile as described in \cite{Gray1992}.  The templates were spun up from 10 $\mathrm{km}\,\mathrm{s}^{-1}$ to 150 $\mathrm{km}\,\mathrm{s}^{-1}$ in increments of 5 $\mathrm{km}\,\mathrm{s}^{-1}$, to cover the expected range of observed $v$~sin~$i$ values.  To ensure the low-velocity end of the calibration curve was well-constrained, we also created broadened spectra with velocities of 3, 5, and 7 $\mathrm{km}\,\mathrm{s}^{-1}$.  The limb-darkening coefficient for each narrow-lined template was estimated with the Exoplanet Characterization Toolkit, using the Kurucz ATLAS9 model grid and estimating solar log(g) and metallicity \citep{Bourque2021}.  We then cross-correlated each broadened spectrum against its original narrow-lined spectrum and measured the FWHM of the peak.  The relationship between each FWHM and its corresponding $v$~sin~$i$ was calibrated by fitting a fourth-order polynomial.  We consider 150 $\mathrm{km}\,\mathrm{s}^{-1}$ to be the highest $v$~sin~$i$ we can reliably measure; above this velocity, the CCF peaks become difficult to fit and the relationship between FWHM and $v$~sin~$i$ is not as consistent.  There was no measurable difference in the FWHM measured for the 3 $\mathrm{km}\,\mathrm{s}^{-1}$ spectra and the 5 $\mathrm{km}\,\mathrm{s}^{-1}$ spectra, so the former was excluded from fitting in order to avoid influencing the low-velocity end of the calibration curve. 

The failure to recover a unique FWHM below 5 $\mathrm{km}\,\mathrm{s}^{-1}$ illustrates the absolute lower limit on a $v$~sin~$i$ with this analysis method, which is related to the dispersion of the spectrograph set-up.  However, the true lower limit for $v$~sin~$i$ is largely dependent on the size of the slit image.  The rotational broadening of a spectral line can only be measured once the intrinsic width of the line is larger than the slit width, which is usually focused to 2 pixels; this is equivalent to 13.6 $\mathrm{km}\,\mathrm{s}^{-1}$ with our spectrograph set-up.  In the middle of the CCD chip, the slit width varied between 1.5 and 1.7 pixels; for spectral lines in these areas, broadening associated with a lower $v$~sin~$i$ may be measurable.  Due to this, we estimate our resolution limit for $v$~sin~$i$ to be 11 $\mathrm{km}\,\mathrm{s}^{-1}$ (best focus between 1.6 and 1.7 pixels), with the caveat that this represents a best case scenario and that reported values near this limit may be upper limits rather than true measurements of $v$~sin~$i$.

The $R_{TD}$ parameter from \cite{Tonry1979} quantifies the signal-to-noise ratio of the cross-correlation peak.  It is proportional to the error of the rotational velocity as $(1 + R_{TD})^{-1}$ \citep{Tonry1979}.  \cite{Hartmann1986} analyzed $v$~sin~$i$ error as a function of $R_{TD}$ for both moderate ($\sim$20 $\mathrm{km}\,\mathrm{s}^{-1}$) and fast ($\sim$50 $\mathrm{km}\,\mathrm{s}^{-1}$) rotating stars and determined that $\pm$ $v$~sin~$i$$/(1 + R_{TD})$ provided a 90\% confidence level for their rotational velocity measurements.  We have implemented this relationship for our own observations as the $v$~sin~$i$ uncertainty for individual measurements.  When we have more than one measurement for a star, 
the reported $v$~sin~$i$ and uncertainty are the weighted mean and its associated uncertainty.

\begin{figure}
    \centering
    \includegraphics[width=0.5\linewidth]{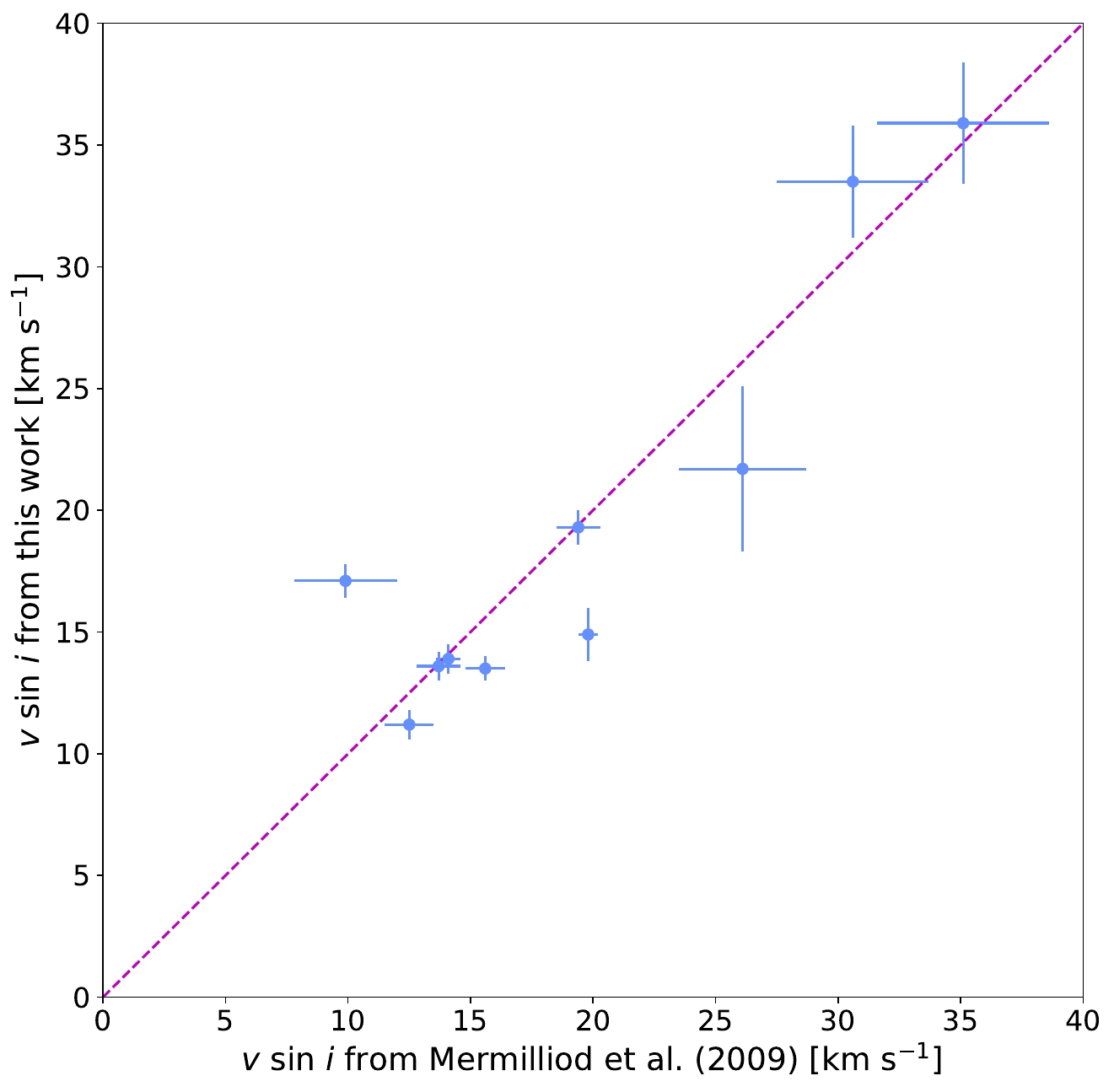}
    \caption{Plot of our measured $v$~sin~$i$ values vs. values for the same stars from \cite{Mermilliod2009}, to demonstrate the effectiveness of our measurement methods.  The magenta dashed line depicts a perfect 1-to-1 correlation.} 
    \label{fig:vsini_standards}
\end{figure}

To investigate how well our measurements correlated with other work at a range of $v$~sin~$i$ values, including those below 13.6 $\mathrm{km}\,\mathrm{s}^{-1}$, we also observed ten bright stars with rotational velocities reported in \cite{Mermilliod2009}, which measured rotation velocities down to about 1 $\mathrm{km}\,\mathrm{s}^{-1}$ with a precision of 1 $\mathrm{km}\,\mathrm{s}^{-1}$.  We measured radial velocity and $v$~sin~$i$ for these stars by cross-correlating their spectra against the narrow-lined template with the closest temperature; in each case, the template star with the closest temperature was the Sun.  The properties of these stars and our measured values are listed in Table \ref{tab:vsini_standards}.  A comparison of our $v$~sin~$i$ measurements to those from \cite{Mermilliod2009} is shown in Fig. \ref{fig:vsini_standards}. We also statistically compared the two sets of measurements using least squares linear regression and the Pearson correlation coefficient $r$.  For two sets of measurements of the same objects, we would expect a slope in agreement with 1 and a correlation coefficient $>$ 0.95.  For all ten stars, $r$ = 0.916, with a slope of 0.94 $\pm$ 0.15.  The reported $v$~sin~$i$ for He 347 is 9.9 $\mathrm{km}\,\mathrm{s}^{-1}$, which is below our resolution limit and so we would not expect to be able to recover a valid $v$~sin~$i$ measurement.  When we exclude He 347, we recover a slope of 1.09 $\pm$ 0.11, with $r$ = 0.965.  Of the nine remaining objects, we measured eight to have $v$~sin~$i$ within two standard deviations of the value reported in \cite{Mermilliod2009}.  This indicates that our methodology is sound and supports the idea that our practical resolution limit can be below 13.6 $\mathrm{km}\,\mathrm{s}^{-1}$. It also demonstrates that our calibration curve is appropriate for the range of $v$~sin~$i$ values covered by the majority of our data (only 8\% of our final sample had $v$~sin~$i$ $>$ 35 $\mathrm{km}\,\mathrm{s}^{-1}$).  

\begin{deluxetable}{ccccccccc}
\tablecaption{Stars from \cite{Mermilliod2009} used for $v$~sin~$i$ measurement comparison}
\tablehead{\colhead{Star} & \colhead{RA} & \colhead{Dec} & \colhead{V\tablenotemark{a}} & \colhead{RV\tablenotemark{b}} & \colhead{RV\tablenotemark{c}} & \colhead{vsin(i)\tablenotemark{b}} & \colhead{vsin(i)\tablenotemark{c}} & \\ 
\colhead{} & \colhead{(deg)} & \colhead{(deg)} & \colhead{(mag)} & \colhead{($\mathrm{km}\,\mathrm{s}^{-1}$)} & \colhead{($\mathrm{km}\,\mathrm{s}^{-1}$)} & \colhead{($\mathrm{km}\,\mathrm{s}^{-1}$)} & \colhead{($\mathrm{km}\,\mathrm{s}^{-1}$)} } 

\startdata
He 347 & 49.40617 & 48.10798 & 10.55 & 8.67 $\pm$ 0.36 & 7.78 $\pm$ 1.11 & 9.9 $\pm$ 2.1 & 17.1 $\pm$ 0.7 \\
AK 1159 & 130.71805 & 17.1337 & 8.76 & 20.5 $\pm$ 0.31 & 19.24 $\pm$ 1.3 & 12.5 $\pm$ 1 & 11.2 $\pm$ 0.6 \\
VL 1353 & 130.65319 & 18.3888 & 10.08 & 35.99 $\pm$ 0.47 & 36.51 $\pm$ 1.2 & 13.7 $\pm$ 0.9 & 13.6 $\pm$ 0.6 \\
Malm 30.252 & 188.75103 & 30.19252 & 8.62 & -1.91 $\pm$ 0.2 & -2.9 $\pm$ 1.08 & 14.1 $\pm$ 0.5 & 13.9 $\pm$ 0.6 \\
He 299 & 48.99553 & 50.4051 & 11.15 & -1.82 $\pm$ 0.48 & -2.78 $\pm$ 0.87 & 15.6 $\pm$ 0.8 & 13.5 $\pm$ 0.5 \\
He 334 & 49.24788 & 49.9265 & 10.37 & -2.47 $\pm$ 0.72 & -3.63 $\pm$ 1.17 & 19.4 $\pm$ 0.9 & 19.3 $\pm$ 0.7 \\
Tr 19 & 183.10367 & 27.38006 & 8.06 & 0.66 $\pm$ 0.24 & -1.85 $\pm$ 2.06 & 19.8 $\pm$ 0.4 & 14.9 $\pm$ 1.1 \\
Tr 101 & 185.92081 & 26.9799 & 8.36 & -0.17 $\pm$ 0.4 & -0.25 $\pm$ 5.14 & 26.1 $\pm$ 2.6 & 21.7 $\pm$ 3.4 \\
KW 411 & 130.40066 & 19.14259 & 9.32 & 35.6 $\pm$ 1.42 & 32.11 $\pm$ 3.06 & 30.6 $\pm$ 3.1 & 33.5 $\pm$ 2.3 \\
Tr 36 & 184.03484 & 25.76033 & 8.07 & 0.03 $\pm$ 0.79 & 0.31 $\pm$ 3.2 & 35.1 $\pm$ 3.5 & 35.9 $\pm$ 2.5 \\
\enddata

\label{tab:vsini_standards}
\tablecomments{Right ascension and declination are given in degrees, using coordinates from Gaia DR3 \citep{GaiaEDR3-2021a}.}
\tablenotetext{a}{V-band magnitude is converted from Gaia DR3 G-band and $G_{BP}-G_{RP}$, using the conversions from \cite{GaiaEDR3Conv2021}.}
\tablenotetext{b}{\cite{Mermilliod2009}}
\tablenotetext{c}{This work.}
\end{deluxetable}

\subsection{Finalizing Measurements and Membership} \label{sec:cleaningdata}

Investigating the rotational properties of PMS stars within NGC 2264 requires good quality measurements of cluster members.  We built our catalog of potential targets from stars that had been determined to be cluster members in one or more of the studies published in the past, but differences in selection criteria across those studies mean that we must define cluster membership for this study and evaluate each object to ensure that we are only including members in our analysis.  Quantitative and qualitative assessments of the FXCOR outputs are necessary to evaluate the quality of the measurements.

While completing the cross-correlations, we took extensive notes about the shape of each peak.  Peaks with structure other than that of a Gaussian function may indicate that a star is a spectroscopic binary manifesting as two (or more) Gaussians overlapping each other, though we note that wider peaks naturally tended to become less Gaussian in shape.  We considered stars that showed some minor structure in the CCF peak, but had no other signs of binarity, to be ``possible binaries". We marked stars that showed a clear double peak in the CCF peak, as well as stars classified as binaries in \cite{Furesz2006}, as ``likely binaries".

We can use the \cite{Tonry1979} $R_{TD}$ to evaluate whether a peak near the expected radial velocity for the cluster is statistically significant.  The noisier a cross-correlation function is, the lower $R_{TD}$ is.  \cite{Furesz2006} conducted a kinematic study of NGC 2264 by measuring radial velocities and indicated that $R_{TD}$ $>$ 2 is an appropriate minimum cut-off for a statistically significant peak, and so we removed any measurements with $R_{TD}$ $\leq$ 2. 

Once we had removed the statistically poor measurements, we went through the measurements for objects that had been observed multiple times.  We specifically looked for cases where measurements were good quality, but had significantly different radial velocities or $v$~sin~$i$; these were also marked as ``likely binaries".  We considered a significant difference to be exceeding 2$\sigma$ disagreement between measurements, or 1$\sigma$ disagreement with noted asymmetry in the CCF.  We included radial velocity measurements from \cite{Furesz2006} and \cite{Jackson2016} in this determination when available.  We also reviewed the cross-correlation function for all of these cases and confirmed that there was no second peak nearby that hadn't been noted, or if there was, that the same peak was measured both times.

Next, we calculated a weighted average for the radial velocity and the $v$~sin~$i$ for all objects with more than one observation.  For objects that had been marked as ``likely binaries" because of discrepant RV or $v$~sin~$i$ measurements, we did not average the discrepant measurements.  For stars with known rotation periods ($P$), luminosities ($L_*$), and temperatures ($T_{\rm eff}$), we can calculate the radius $R_*$ = $\sqrt{L_*/4 \pi \sigma_{SB} T_{eff}^4}$.  Then, we calculated the equatorial rotation velocity, $V_{eq}$ = 2$\pi R_*/P$.  Any star that had a measurement of $v$~sin~$i$ or estimated $V_{eq}$ $<$ 11 $\mathrm{km}\,\mathrm{s}^{-1}$ was marked as being an upper limit measurement.  One star, 1136, had $v$~sin~$i$ measured to be $>$ 150 $\mathrm{km}\,\mathrm{s}^{-1}$ and was not included in further analysis because this was outside of the range where there is good correlation between the FWHM and $v$~sin~$i$.

We used the radial velocity to determine membership in the cluster.  \cite{Furesz2006} defined membership in NGC 2264 with a radial velocity range of 8$-$36 $\mathrm{km}\,\mathrm{s}^{-1}$.  Objects with RVs in agreement with this range were considered to be cluster members. The radial motions of binary systems can complicate this process, as one or more components may have a radial velocity outside of the accepted range.  For simplicity, any suspected binary with at least one radial velocity measurement inside the expected range was considered to be a cluster member.  When the midpoint of two RV measurements (from a double peak or from two different observations of a single peak) was within the range, even if neither RV measurement was, we also considered that object to be a member.  Excluding binary objects, the average RV of our Main Analysis Sample (see Section \ref{sec:subsamples} for description) was 20.32 $\mathrm{km}\,\mathrm{s}^{-1}$ with a velocity dispersion of 4.05 $\mathrm{km}\,\mathrm{s}^{-1}$.

\subsection{Statistical Comparison of Projected Stellar Radii to Radii Predicted by Stellar Evolutionary Models} \label{sec:maxlikelihoodmethod}

It has been established that when direct radius measurements of low-mass PMS eclipsing binaries are compared to radius predictions from PMS evolutionary models, the stellar radii are found to be larger than expected; this discrepancy has been termed ``radius inflation" (e.g., \citealt{Lopez-Morales2007, Torres2010, Kraus2015, David2019, Smith2021}).  Statistical modeling techniques such as those employed by \cite{Lanzafame2017} and \cite{Jackson2018} use the projected stellar radius, calculated from the rotation period and the projected rotation velocity, to estimate the average stellar radius for a cluster of stars.  Those works have found evidence of radius inflation in the 120 Myr old Pleiades cluster.  We will use similar techniques to investigate the accuracy of model radius predictions for the much younger NGC 2264 cluster. 

\cite{Jackson2018} used a maximum likelihood method to demonstrate that M dwarfs in the Pleiades cluster had radii that were 14\% $\pm$ 2\% larger than the predictions from evolutionary models, and we have adopted their approach.  A full explanation of the maximum likelihood method is included in Section 4 of \cite{Jackson2018}, but can be summarized as follows: For a star with a measured rotation period $P$ and projected rotation velocity $v$~sin~$i$, the projected stellar radius can be calculated as $R$~sin~$i$ = $P$ $v$~sin~$i$/(2$\pi$) (e.g., \citealt{Rhode2001b}).  This value also represents a lower limit on the radius of the star.  We define $r$~sin~$i$ = $R$~sin~$i$/$R_m$, the ratio of the projected radius to the model radius for a star of the same luminosity, $R_m$.  We also define the the average radius ratio of the measured radii to the predicted model radii for the group of objects, $\rho$ = R/$R_m$. Assuming a random spin axis orientation distribution \citep{Jackson2010a} and a given value of $\rho$, we use Monte Carlo methods to build a probability distribution of possible values for $r$~sin~$i$ where $i$ is drawn from a random uniform distribution of 0$^{\circ}$ to 90$^{\circ}$.  

\begin{figure}
    \centering
    \includegraphics[width=0.5\linewidth]{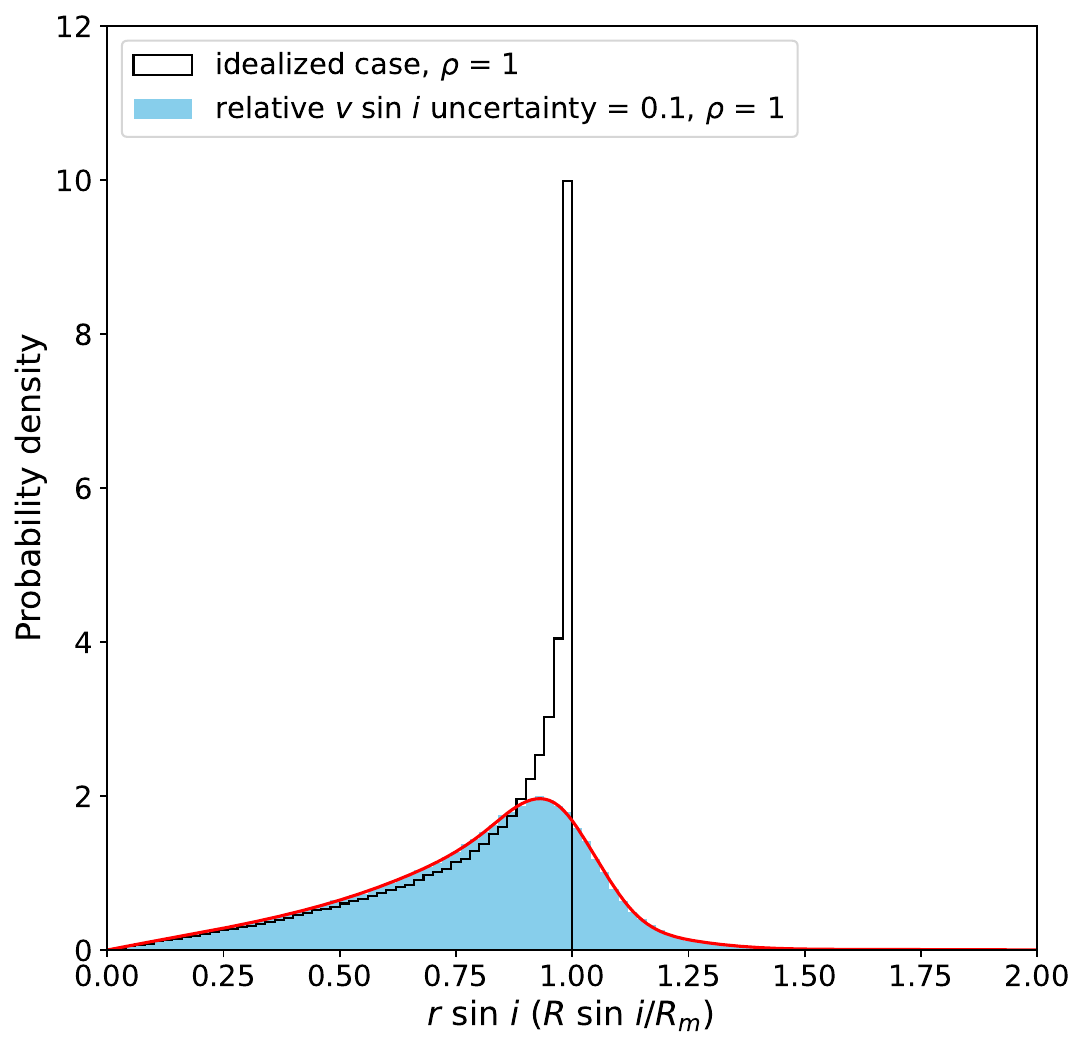}
    \caption{Example probability density distribution of $r$~sin~$i$, for $\rho$ = 1.  The black histogram is for the idealized case, with no measurement uncertainties in $v$~sin~$i$ or the rotation period. The blue histogram is the probability distribution for a star with a relative uncertainty of 0.1 in the $v$~sin~$i$ measurement, with the red line showing the curve fit to the distribution.  The probability of measuring a specific value of $r$~sin~$i$ given the rotation period, $v$~sin~$i$, and radius predicted by the stellar evolutionary model can then be obtained from the fitted function.}
    \label{fig:pdf_example}
\end{figure}

The probability distribution will be broadened by the relative uncertainty in $r$~sin~$i$, which is proportional to the relative uncertainty in the $v$~sin~$i$ measurement.  For each value of $r$~sin~$i$ that was drawn to build the probability distribution, we also draw a random value from a Student's t-distribution with $\nu$ = 3. This second value is multiplied by the relative uncertainty in $v$~sin~$i$ to become the relative uncertainty in $r$~sin~$i$.  It is then multiplied by the associated $r$~sin~$i$ to become the absolute uncertainty, and added to $r$~sin~$i$ before it is included in the probability distribution.  For values of $v$~sin~$i$ $<$ 11 $\mathrm{km}\,\mathrm{s}^{-1}$, we do not know the true value of $v$~sin~$i$ and its associated uncertainty, only that it is between 0 and 11 $\mathrm{km}\,\mathrm{s}^{-1}$.  To build an appropriate probability distribution for these objects, we consider them to have $v$~sin~$i$ = 5.5 $\pm$ 5.5 $\mathrm{km}\,\mathrm{s}^{-1}$; effectively, a relative uncertainty of 1.  An example probability distribution, using a $v$~sin~$i$ relative uncertainty of 0.1, for the case where $\rho$ = 1, is shown in Fig. \ref{fig:pdf_example}.

For a range of values of $\rho$, a probability distribution is calculated for each object in the sample and we fit a curve to the distribution to calculate the probability of observing a specific value of $r$~sin~$i$ given the object's period, $v$~sin~$i$, and radius prediction from the model.  For objects with $v$~sin~$i$ $<$ 11 $\mathrm{km}\,\mathrm{s}^{-1}$, the probability distribution function (PDF) is integrated between 0 and $r$~sin~$i_{UL}$, the upper limit of $r$~sin~$i$ from the object's period and $v$~sin~$i$ = 11 $\mathrm{km}\,\mathrm{s}^{-1}$, and the result is divided by $r$~sin~$i_{UL}$, resulting in the average value of the PDF between 0 and $r$~sin~$i_{UL}$ (see \citealp{Jackson2018}, Eq. 6). 

The log-likelihood function is the sum of the natural log of the probabilities of each object for a given value of $\rho$ (see \citealp{Jackson2018}, Eq. 7).  The best-fitting value of $\rho$ is at the maximum of the log-likelihood function, ln $\widehat{\mathcal{L}}$, and the uncertainty of $\rho$ is obtained from the standard deviation of the likelihood function.

The model radius for a star was defined as the radius predicted by the model for a star of equivalent luminosity.  Different models cover different ranges in luminosity, so to be able to use the same sample of objects for each model, we fitted a polynomial function to the luminosity-radius relationship for a given isochrone and starspot fraction and extrapolated it for objects that extended beyond the maximum or minimum of the model range, though these do not represent a significant portion of the sample.  

While \cite{Jackson2018} divided objects into three subgroups based on luminosity to investigate a possible mass dependence, we divided our sample based on temperature instead.  The low-mass objects in NGC 2264 are still on the Hayashi track, where the mass tracks are nearly vertical on an HR diagram, so temperature is a more appropriate proxy for mass than luminosity in this era \citep{Hayashi1961}. The three temperature bins were created to have as close to equal numbers of objects in each group as possible, and are defined in Section \ref{sec:model-comp}.

\section{Results \& Discussion} \label{sec:results}

\subsection{Samples and Sub-samples for Analysis} \label{sec:subsamples}

Our final sample is comprised of 254 low-mass PMS stars in NGC 2264, listed in Table \ref{tab:targets}.  There are 176 ``single systems", which showed no signs of binarity (i.e., they showed symmetrical CCF peaks and, if observed multiple times, had consistent RV and $v$~sin~$i$ values).  There are an additional 34 ``possible binaries", which had at least one CCF peak with a small asymmetry, but no other indications of binarity such as disagreement in RV measurements from different epochs.  Finally, there are 44 ``likely binaries", which are objects that showed multiple CCF peaks, had varying RV or $v$~sin~$i$ measurements, or had been identified as a binary system by \cite{Furesz2006}.  A full explanation of these classifications is in Section \ref{sec:cleaningdata}.  The first two groups (single systems and possible binaries) comprise the Main Analysis Sample.  This is further broken into two subsets.  Both subsets exclude the object for which we measured $v$~sin~$i$ $>$ 150 $\mathrm{km}\,\mathrm{s}^{-1}$. The first subset, which will be referred to as the Resolution-Limited Sample, consists of the 112 objects which have $v$~sin~$i$ and $V_{eq}$ greater than our resolution limit of 11 $\mathrm{km}\,\mathrm{s}^{-1}$ and is mainly used in the inclination and $v$~sin~$i$ distribution analyses of Sections \ref{sec:inc} and \ref{sec:disks}.  The second subset, the Model Comparison Sample, is used in Section \ref{sec:model-comp} and contains 136 objects.  This subset is limited to objects with a luminosity and a period measurement from the literature, which is required for the statistical maximum likelihood method described in Section \ref{sec:maxlikelihoodmethod} (luminosity to select an appropriate model radius, and period to estimate a projected stellar radius).  Stars with $v$~sin~$i$ $<$ 11 $\mathrm{km}\,\mathrm{s}^{-1}$ are permitted in the Model Comparison Sample, as the maximum likelihood method allows an upper limit for $r$~sin~$i$ to be considered, but stars with $V_{eq}$ $<$ 11 $\mathrm{km}\,\mathrm{s}^{-1}$ are still excluded as they would not contribute to constraining the average radius ratio.  There are 101 objects included in both subsets, with the main divergence coming from the first subset's exclusion of objects with $v$~sin~$i$ below the velocity resolution limit.

\begin{deluxetable}{ccccccccccccc}[h]
\tablewidth{1.\columnwidth}
\tabletypesize{\footnotesize}
\tablecaption{Measured values of RV and $v$~sin~$i$ for stars in NGC 2264, with auxiliary data adopted from the literature}
\label{tab:targets}
\tablehead{\colhead{ID} & \colhead{Gaia DR3 ID} & \colhead{RA} & \colhead{Dec} & \colhead{$T_{\rm eff}$\tablenotemark{a}} & \colhead{Per\tablenotemark{a}} & \colhead{$L_{\rm bol}$\tablenotemark{a}} & \colhead{Rad\tablenotemark{a}} & \colhead{$V_{eq}$\tablenotemark{b}} & \colhead{Disk\tablenotemark{a}} & \colhead{Binarity\tablenotemark{c}} & \colhead{RV} & \colhead{$v$~sin~$i$} \\ 
\colhead{} & \colhead{} & \colhead{(deg)} & \colhead{(deg)} & \colhead{(K)} & \colhead{(days)} & \colhead{($L_{\odot}$)} & \colhead{($R_{\odot}$)} & \colhead{($\mathrm{km}\,\mathrm{s}^{-1}$)} & \colhead{} & \colhead{} & \colhead{($\mathrm{km}\,\mathrm{s}^{-1}$)} & \colhead{($\mathrm{km}\,\mathrm{s}^{-1}$)} }

\startdata
1 & G3326685443313414144 & 100.31563 & 9.43801 & 4140 & 6.227 & 3.6 & 3.688 & 29.97 & W & A & 21.96 $\pm$ 2.02 & 20.4 $\pm$ 1.1 \\
3 & G3326710968303264384 & 100.28235 & 9.68747 & 3770 & 1.963 & 0.75 & 2.030 & 52.32 & W & A & 26.69 $\pm$ 6.15 & $<$11  \\
4 & G3326693101239225472 & 100.35188 & 9.54586 & 2880 & 1.048 & 0.177 & 1.690 & 81.58 & W & A & 23.92 $\pm$ 3.84 & 16.4 $\pm$ 1.8 \\
5 & G3326717019912569088 & 100.19792 & 9.82469 & 4550 & 4.3 & 3.9 & 3.178 & 37.39 & C & A & 22.81 $\pm$ 1.48 & 21.7 $\pm$ 0.8 \\
7 & G3326715439364610560 & 100.18767 & 9.76160 & 3970 & 4.612 & 0.57 & 1.596 & 17.51 & W & b & 20.44 $\pm$ 3.03 & 24.0 $\pm$ 1.8 \\
8 & G3326702004706943872 & 99.99934 & 9.56156 & 3970 & 2.575 & 0.32 & 1.196 & 23.49 & W & A & 15.61 $\pm$ 2.41 & 23.5 $\pm$ 1.4 \\
9 & G3326739766059353472 & 100.28953 & 9.86387 & 3360 & 5.956 & 1.01 & 2.966 & 25.19 & C & B & -19.98 $\pm$ 4.35 & 15.7 $\pm$ 2.0 \\
10 & G3326689291604269312 & 100.18094 & 9.47802 & 3770 & 9.25 & 0.57 & 1.770 & 9.68 & C & A & 16.75 $\pm$ 1.05 & $<$11  \\
12 & G3326716092199626880 & 100.27123 & 9.81331 & 3770 & 3.756 & 0.78 & 2.070 & 27.89 & W & A & 21.96 $\pm$ 1.11 & 18.2 $\pm$ 0.6 \\
16 & G3326719047137454848 & 100.48304 & 9.67967 & 3425 & 1.225 & 0.52 & 2.048 & 84.58 & W & A & 18.33 $\pm$ 2.49 & $<$11 \\
\enddata

\tablecomments{Right ascension and declination are given in degrees, using coordinates from Gaia DR3 \citep{GaiaEDR3-2021a}.  A sample of the table is shown here, the full version is available online.}
\tablenotetext{a}{Compiled from the literature.  See Section \ref{sec:adoption} for a detailed discussion of how $T_{\rm eff}$, period, $L_{\rm bol}$, radius, and disk classification were selected.}
\tablenotetext{b}{Calculated from the adopted radius and period.}
\tablenotetext{c}{``A" indicates a star that has been classified as a single star system, ``b" indicates a ``possible" binary system, and ``B" indicates a ``likely" binary system.  For a full discussion on classifications, see Section \ref{sec:cleaningdata}.}

\end{deluxetable}

\subsection{Comparing Our $v$~sin~$i$ Measurements With Other Studies of NGC 2264}

\begin{figure}
    \centering
    \includegraphics[width=0.9\linewidth]{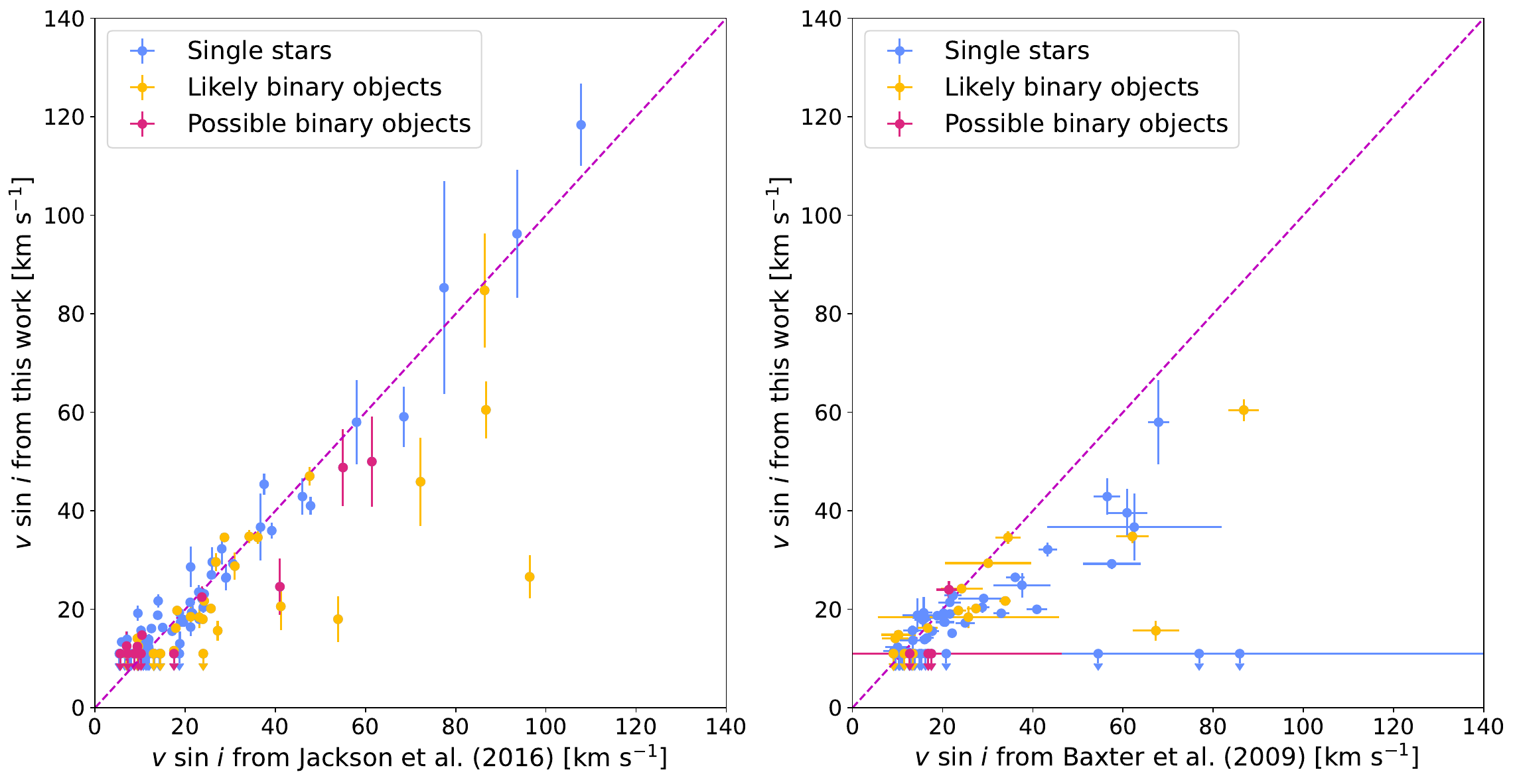}
    \caption{Comparison of our $v$~sin~$i$ measurements with measurements from \cite{Jackson2016} (left) and \cite{Baxter2009} (right).  The error bars on the measurements from \cite{Jackson2016} are smaller than the markers.  Objects with $v$~sin~$i$ $<$ 11 $\mathrm{km}\,\mathrm{s}^{-1}$ are marked as upper limits with arrows.  Objects we identified as ``likely" binary systems are marked in yellow, while ``possible" binaries are in pink and ``single" stars are in blue.  The magenta dashed line depicts a perfect 1-to-1 correlation.  We find good agreement with the measurements from \cite{Jackson2016}, but not with \cite{Baxter2009}.}
    \label{fig:vsini_vs_both}
\end{figure}

To check the accuracy of our methodology, we compare our $v$~sin~$i$ measurements to those of \cite{Jackson2016} and \cite{Baxter2009}.  The left panel of Figure \ref{fig:vsini_vs_both} shows a comparison between our $v$~sin~$i$ values and \cite{Jackson2016}, for 117 objects in common between the two samples.  This comparison excludes 23 objects for which \cite{Jackson2016} reported velocities below their resolution limit of 5 $\mathrm{km}\,\mathrm{s}^{-1}$.  We find good agreement with the measurements of \cite{Jackson2016}, with most exceptions able to be explained as binaries; these are likely cases where the CCF may have been overly widened depending on the orientation of the binary system at the time of observation, or the measurements may be of different components.  We want to compare our measurements without any of the complications that binaries may introduce, so we do a linear least squares regression only using objects identified as single systems with $v$~sin~$i$ and $V_{eq}$ $\geq$ 11 $\mathrm{km}\,\mathrm{s}^{-1}$ (50 objects).  This gives a slope of 1.00 $\pm$ 0.03, with an intercept of 0.89 $\pm$ 1.00 and Pearson r = 0.980; an excellent correlation with no sign of systematic error in the measurement.   

The comparison to \cite{Baxter2009} is less straightforward.  There are 71 objects in common between our Main Analysis Sample and the sample from \cite{Baxter2009}.  As shown in the right panel of Fig. \ref{fig:vsini_vs_both}, there are significant differences between the $v$~sin~$i$ measurements for several single stars, particularly for objects that \cite{Baxter2009} measured with $v$~sin~$i$ above $\sim$25 $\mathrm{km}\,\mathrm{s}^{-1}$.  There are seven single stars in common with \cite{Jackson2016} that \cite{Baxter2009} measured $v$~sin~$i$ $>$ 25 $\mathrm{km}\,\mathrm{s}^{-1}$; six of the measurements from \cite{Jackson2016} are in agreement with our measurements.  Performing a linear least squares regression on the 34 single stars with $v$~sin~$i$ and $V_{eq}$ $\geq$ 11 $\mathrm{km}\,\mathrm{s}^{-1}$ gives a slope of 0.55 $\pm$ 0.04 and an intercept of 6.56 $\pm$ 1.40, a poor agreement between measurements.  The reasons for the discrepancies between our measurements and those of \cite{Baxter2009} are unclear, but the fact that our data agree well with the \cite{Jackson2016} values over our full $v$~sin~$i$ range lends confidence to our results.

\subsection{The Inclination Distribution of Our Sample} \label{sec:inc}

\begin{figure}
    \centering
    \includegraphics[width=0.5\linewidth]{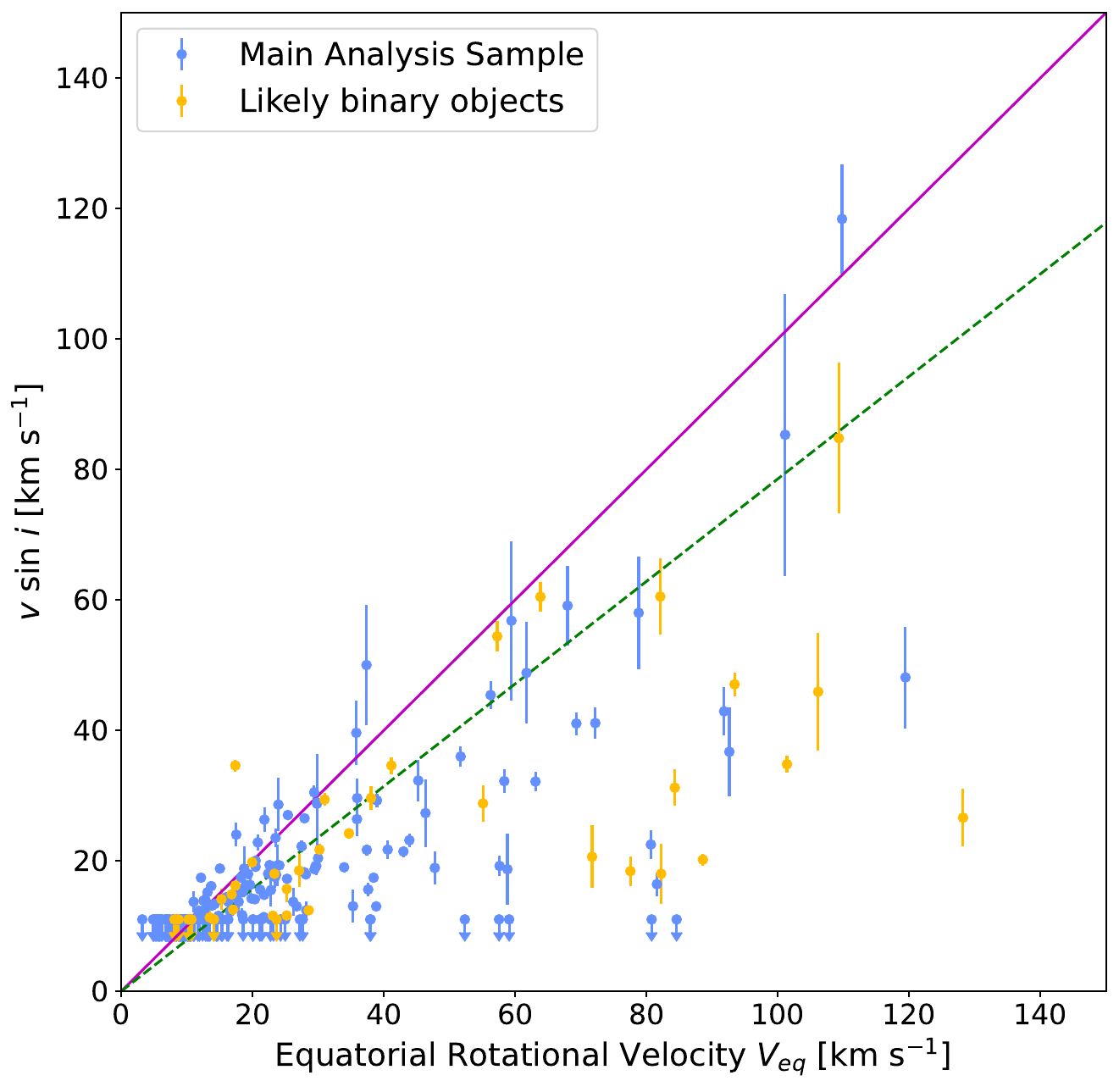}
    \caption{Measured $v$~sin~$i$ vs. equatorial rotation velocity ($V_{eq}$) for the 229 low-mass stars in NGC 2264 for which both were measured.  Single stars and ``possible" binaries are shown in blue and comprise the Main Analysis Sample.  Objects we identified as ``likely" binary systems are marked in yellow, and objects with $v$~sin~$i$ $<$ 11 $\mathrm{km}\,\mathrm{s}^{-1}$ are marked as upper limits with arrows. The magenta line marks $v$~sin~$i$ = $V_{eq}$ and the dashed green line marks $v$~sin~$i$ = 0.785 $V_{eq}$.  A randomly-distributed sample of stellar spin-axes is expected to have $\langle$sin $i$$\rangle$ = 0.785.  For the Resolution-Limited Sample, we calculate $\langle$sin $i$$\rangle$ = 0.784 $\pm$ 0.025; this indicates that our sample of stars appears to have randomly distributed inclinations and we don't identify any systematic errors for our values adopted from the literature or our $v$~sin~$i$ measurements.}
    \label{fig:vsini_vs_veq}
\end{figure}

\cite{Healy2023} studied the inclinations of 11 open clusters, and found that eight had spin-axis orientations that were consistent with isotropy (although they were unable to completely rule out the possibility of moderate alignment).  Additionally, \cite{Jackson2010a} and \cite{Jackson2018} found no strong evidence for alignment in the Pleiades or Alpha Per clusters. \cite{Kovacs2018} and \cite{Corsaro2017} have suggested that the Praesepe, NGC 6791, and NGC 6819 clusters may show alignment, but those results have been refuted by \cite{Jackson2019} and \cite{Mosser2018}.  \cite{Aizawa2020} explored whether the position angles of circumstellar disks were aligned in five regions and found only marginal support for alignment in Lupus III; while disks may not be perfectly aligned to stellar spin-axes, they do act as a proxy for general alignment trends in a cluster.  The expected mean $\langle$sin $i$$\rangle$ for a cluster with randomly distributed spin-axes is 0.785 \citep{Chandrasekhar1950}.  As explored in \cite{Rhode2001b}, deviations from this value could indicate systematic errors in the measurement of $v$~sin~$i$, the rotation period, or estimation of the temperature or luminosity (from uncertainties in the distance, extinction correction, or the inclusion of unresolved binary companions). For NGC 2264, \cite{Rebull2004} estimated $\langle$sin $i$$\rangle$ = 0.85 $\pm$ 0.12, based on a sample of seven stars.  Fig. \ref{fig:vsini_vs_veq} shows a comparison between the measured $v$~sin~$i$ values and $V_{eq}$ for both the objects in the Main Analysis Sample and the ``likely" binary objects.  Using our Resolution-Limited Sample, we calculate $\langle$sin $i$$\rangle$ = 0.784 $\pm$ 0.025.  Therefore, we assume a random distribution of spin-axes for our analyses of NGC 2264, and our accepted values for the periods, temperatures, and luminosities appear to be appropriate.

\subsection{$v$~sin~$i$ Distributions}

\subsubsection{Examining the Effects of Accreting Disks on $v$~sin~$i$} \label{sec:disks}

Many studies have investigated whether circumstellar disks may influence the $v$~sin~$i$ distribution of young stars in a cluster. In some cases, investigators found no clear correlation between disk markers and rotation rate (e.g., \citealt{Stassun1999, Rebull2001, Rebull2004, Makidon2004, Littlefair2010}), while others found that stars with disk signatures have average rotation periods that are slower than stars without such signatures (e.g., \citealt{Edwards1993, Herbst2001, Herbst2002, Lamm2005, Cieza2007b, Venuti2017, Rebull2018}).  In a study of T Tauri star rotation in the Orion star-forming complex, \cite{Serna2021} found that in the first $\sim$5-6 Myr, despite continued contraction of the radius, $v$~sin~$i$ tended to decrease overall for stars.  The observed $v$~sin~$i$ trend followed predictions from a set of \cite{Gallet2015} evolutionary models which included disk interaction.  The end of this period, when the average $v$~sin~$i$ of the stars began to increase, coincided with higher estimates for the time by which most disks are expected to have dissipated \citep{Fedele2010, Gallet2015, Pfalzner2022}.  \cite{Serna2021} proposed disk-locking as the most significant process removing angular momentum from the stars during this time. \cite{Serna2024} investigated the rotation evolution of CTTSs with a set of stellar rotation models incorporating accretion-powered stellar winds, whereby some fraction of the mass being accreted from the disk is transferred to a stellar wind, resulting in the loss of angular momentum. This, in addition to interaction between the magnetic field lines and the disk, led to an equilibrium in the rotation rate, demonstrating how the star's rotation rate is moderated by the interaction with the disk. At 3 Myr, NGC 2264 should still have a substantial population of stars with disks, allowing us to do a statistical investigation into the two classes of T Tauri stars.  

In the Main Analysis Sample (i.e. excluding ``likely binaries"), there are 80 CTTSs, 123 WTTSs, and 5 CWTTSs (two objects were excluded because they had no information in the literature that indicated disk status).  A disked population of $\sim$38 $\pm$ 4\% is in agreement with populations of low-mass stars identified as disk-bearing in other 3 Myr old clusters such as $\sigma$ Orionis \citep{Oliveira2006, Hernandez2007}. 

\begin{figure}
    \centering
    \includegraphics[width=0.5\linewidth]{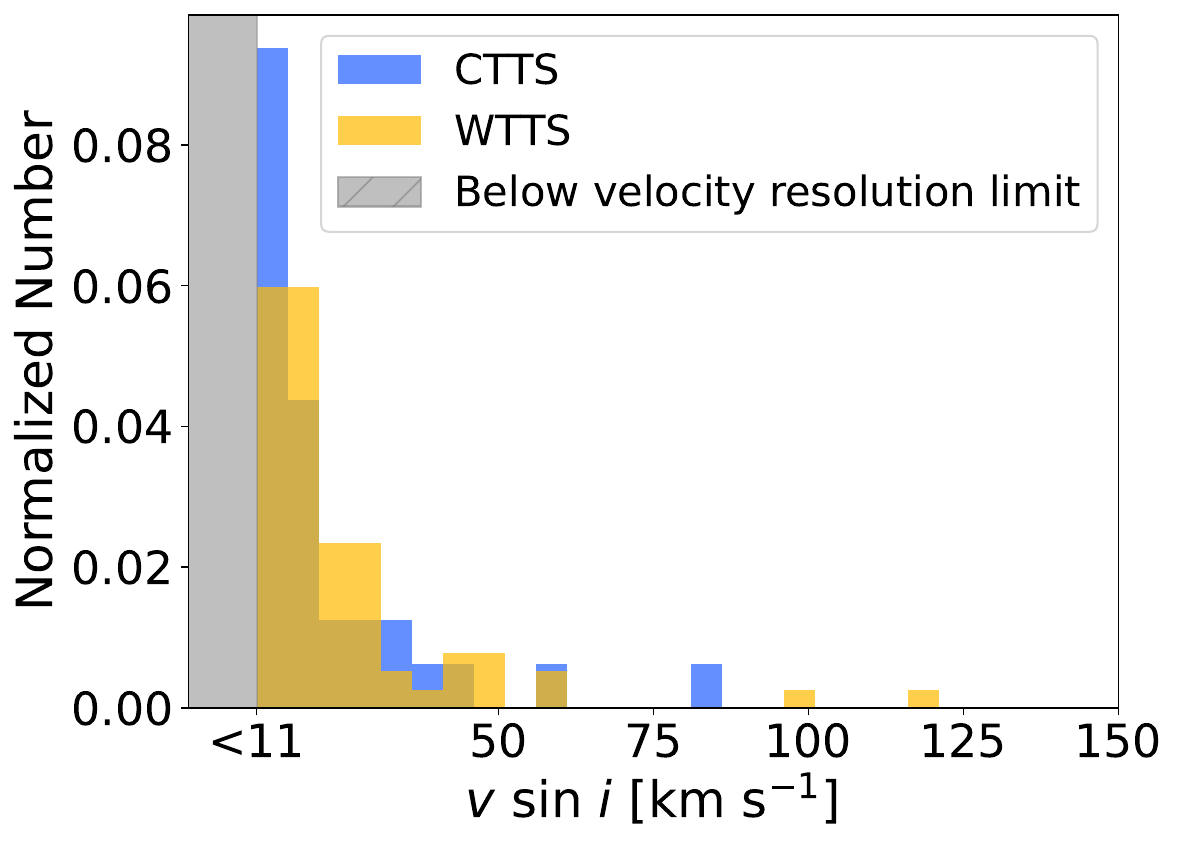}
    \caption{Normalized $v$~sin~$i$ distributions for disked, CTTS stars (blue) and diskless, WTTS stars (yellow). The gray bar marks the area below the velocity resolution limit.  The WTTS stars appear to rotate faster, on average, than the CTTS stars, and very few CTTS stars rotate faster than 50 $\mathrm{km}\,\mathrm{s}^{-1}$.}
    \label{fig:disk_distrib}
\end{figure}

We find that 60 $\pm$ 9\% (48/80) of the CTTSs have $v$~sin~$i$ below 11 $\mathrm{km}\,\mathrm{s}^{-1}$, while only 37 $\pm$ 6\% (46/123) of WTTSs are below this velocity resolution limit; a much larger fraction of CTTSs rotate slowly.  Conversely, we can compare the ``fast rotator" fractions, as \cite{Serna2021} did.  They defined the $v$~sin~$i$ of a ``fast rotator" such that approximately 25\% of their WTTSs exceeded that speed; for our sample, 25\% (30/123) of WTTSs have $v$~sin~$i$ $\geq$ 23.5 $\mathrm{km}\,\mathrm{s}^{-1}$.  \cite{Serna2021} found that only 12\% of CTTS stars in their sample were fast rotators.  Similarly, we find that 11 $\pm$ 4\% (9/80) of our CTTS stars exceed 23.5 $\mathrm{km}\,\mathrm{s}^{-1}$.  The $v$~sin~$i$ distributions of the stars with $v$~sin~$i$ and $V_{eq}$ above 11 $\mathrm{km}\,\mathrm{s}^{-1}$ are shown in Fig. \ref{fig:disk_distrib}.

Our statistical analysis is somewhat limited by the proportion of stars below our resolution limit, but we conduct a Kolmogorov-Smirnov (K-S) test on the CTTS and WTTS populations in the Resolution-Limited Sample and recover a p-value of 0.399, or a 39.9\% chance that both samples are drawn from the same underlying probability distribution.  This result suggests that there is no statistically significant difference between the two populations, but we note that with a large portion of the samples being excluded by the velocity resolution limit, we cannot draw a firm conclusion.  

We originally defined our CTTS sample by considering both $H\alpha$ equivalent widths and/or IR excess; however, it is possible for stars that lack accretion signatures to have an IR excess.  In \cite{Venuti2018}, 25.8\% of stars with IR excesses in their NGC 2264 sample were also classified as ``non-accreting" by their $H\alpha$ equivalent widths.  These stars may simply have been going through a period of low accretion activity at the time of the measurement, as noted in \cite{Venuti2018}, or the accretion phase may have finished while the disk remained present \citep{Hernandez2023}.  To specifically explore a possible connection between disk interaction and stellar rotation, we reclassify objects using extreme criteria: only objects with $H\alpha$ equivalent width above the thresholds defined by \cite{Briceno2019} (indicating active accretion, i.e. star-disk interaction) are classified as ``disk-interacting", and only objects with measurements that never indicated accretion or the presence of a disk (such as IR excess) are classified as ``disk-free".  In this new categorization, there are 23 ``disk-interacting" stars and 75 ``disk-free" stars above the velocity resolution limit.  The K-S test p-value for these $v$~sin~$i$ distributions is 0.087, which is a statistically significant difference at the 10\% level but not the 5\% level.

\cite{Serna2021} found a statistically significant difference between CTTS and WTTS stars (K-S p = 0.03) in the Orion Star Forming Complex.  They used rotational velocities from APOGEE spectra \citep{APOGEE2017}, which have a similar resolution limit to our data of 13 $\mathrm{km}\,\mathrm{s}^{-1}$, though objects below the resolution limit were included in their statistical test.  When the sample from \cite{Serna2021} is limited to objects with $v$~sin~$i$ $\geq$ 13 $\mathrm{km}\,\mathrm{s}^{-1}$, the K-S p = 0.012, which is an even stronger probability that the CTTS and WTTS populations have different velocity distributions.  \cite{Nofi2021}, who measured $v$~sin~$i$ for a group of PMS stars in the Taurus$-$Auriga star-forming region, also found a statistically significant difference in these two populations (K-S p = 0.002), with a much lower velocity resolution limit of 4 $\mathrm{km}\,\mathrm{s}^{-1}$.  However, they stated that they may have introduced a selection bias in constructing their sample, and so could not determine whether there was a true statistical difference \citep{Nofi2021}.

\subsubsection{Examining the Effects of Multiplicity on $v$~sin~$i$}

Measuring $v$~sin~$i$ with the cross-correlation method introduces additional complications for a binary object.  An unresolved companion can cause blending in the absorption lines and artificially broaden the CCF peak, leading to an over-estimated $v$~sin~$i$.  There may be two visible CCF peaks, one from each component, ranging from a distinct secondary bump on the side of the main peak to fully resolved and separated peaks.  Additionally, in the case of multiple observations of a system, the highest CCF peak may not correspond to the same component each time.  For this analysis, we compare the $v$~sin~$i$ distributions of the 176 ``single systems" to the 44 ``likely binaries".  We exclude the 34 ``possible binaries" to provide the clearest distinction between the two groups.  We find that 45 $\pm$ 5\% (79/176) of the single stars in our sample have $v$~sin~$i$ below 11 $\mathrm{km}\,\mathrm{s}^{-1}$, while only 25 $\pm$ 8\% (11/44) of the binary stars have $v$~sin~$i$ below this velocity resolution limit.  

\begin{figure}
    \centering
    \includegraphics[width=0.5\linewidth]{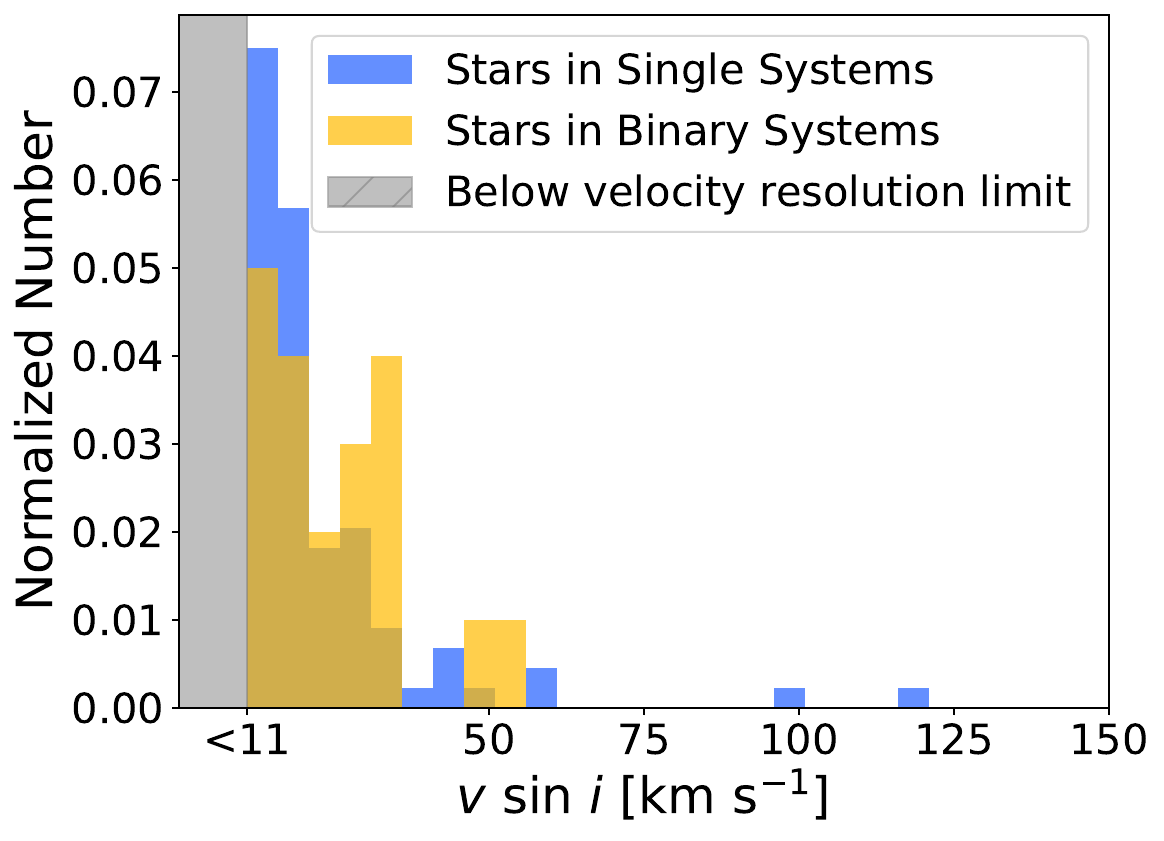}
    \caption{Normalized $v$~sin~$i$ distributions for stars we identified as ``quality single stars" (blue) and ``quality likely binary" systems in our analysis (yellow).  Only objects with $R_{TD}$ $\geq$ 5 are included.  The gray bar marks the area below the velocity resolution limit.  The binary stars appear to rotate faster, on average, than the single stars.}
    \label{fig:binary_distrib}
\end{figure}

As with the analysis in Section \ref{sec:disks}, our statistical analysis is somewhat limited by the proportion of objects with $v$~sin~$i$ below the resolution limit.  Furthermore, given the complications in measuring $v$~sin~$i$ for binary systems, we want to ensure that we are using the most robust set of measurements.  We applied a cut of $R_{TD}$ $\geq$ 5 to both samples of objects with $v$~sin~$i$ $>$ 11 $\mathrm{km}\,\mathrm{s}^{-1}$, to get 88 ``quality single" stars and 20 ``quality likely binary" stars.  This cut was chosen after carefully re-examining the CCF peaks and $v$~sin~$i$ measurements of the ``likely" binary objects to select only those with the most consistent and reliable measurements; the lowest $R_{TD}$ remaining in this group was 5.  As it is possible for the CCF to be widened by a close binary companion, resulting in an over-estimate for the $v$~sin~$i$ measurement, we calculate $\langle$sin $i$$\rangle$ for the group of ``quality likely binaries" to evaluate whether our measurements are over-estimated.  We find $\langle$sin $i$$\rangle$ = 0.708 $\pm$ 0.078, which is in agreement with the expected value of 0.785.  The $v$~sin~$i$ distributions of the stars with $v$~sin~$i$ and $V_{eq}$ above 11 $\mathrm{km}\,\mathrm{s}^{-1}$ and $R_{TD}$ $\geq$ 5 are shown in Fig. \ref{fig:binary_distrib}.  While the binary stars do appear to have a faster average rotation and a wider distribution than the single stars, we hesitate to describe it as bimodal due to the small size of the binary sample.  A K-S test between the ``quality likely binary" and ``quality single" $v$~sin~$i$ distributions resulted in p = 0.112.  When performing a K-S test on the $v$~sin~$i$ distributions of single stars against stars in binary, triple, or quadruple systems, \cite{Nofi2021} found p = 0.056.  The lower velocity resolution limit of their data was about 4 $\mathrm{km}\,\mathrm{s}^{-1}$, which means their distribution was not as limited as ours was.  They also have an upper limit of 50 $\mathrm{km}\,\mathrm{s}^{-1}$, but we did not measure a significant portion of stars above this limit and so it should not affect our ability to compare results.  We found that if a $v$~sin~$i$ cut of 11 $\mathrm{km}\,\mathrm{s}^{-1}$ is applied to the results from \cite{Nofi2021}, so that it resembles the limits of our sample, then the K-S test p-value becomes 0.59.  This lends some weight to the idea that our resolution limit may negatively affect the K-S test's ability to distinguish between two populations in our data.  

The objects in our ``quality likely binary" sample were primarily identified as binaries by differences in RV at different epochs or by the appearance of clear double peaks in the CCF, indicating two sets of lines in the stellar spectrum; therefore, these binaries are quite close to each other.  A few ideas that would explain why close binaries may rotate faster than single stars have been proposed.  The circumstellar disk of a star may be truncated or otherwise disrupted by a close binary companion, which would limit the impact of any angular momentum-draining mechanisms such as disk-locking \citep{Stauffer2018}.  \cite{Bouvier1991} has suggested that a population of short period stars in a bimodal distribution of rotation periods may be explained as close binaries that have been spun up by tidal interactions.  Close binaries can eventually become tidally locked, meaning that their rotation periods are actually the orbital period of the binary system and no longer a consequence of the angular momentum evolution of the star due to contraction (e.g., \citealt{Levato1974, Zahn1989}).  This would make it difficult to compare their velocity distributions to those of single stars, as the dynamics of each binary system would be different.  However, \cite{Fleming2019} indicates that the timescales involved for stars to become tidally locked are $>$ 3 Myr, so the rotation periods for binary stars in NGC 2264 should still reflect the spot-modulated rotation period of the star and not the orbital period of the binary system.  

Using information about the disk status of the stars, we can try to untangle the two potential effects.  In single systems, 40 $\pm$ 5\% (69/174) of stars are CTTSs, while in binary systems, only 32 $\pm$ 9\% (14/44) are CTTSs.  This provides some evidence that the binary systems may be more likely to lack disks than their single counterparts, which may contribute to increased average rotation speeds, but the small sample size of binary CTTSs is far from conclusive.  When we compare the CTTS single and binary stars, 56 $\pm$ 9\% (38/69) of disked single stars have $v$~sin~$i$ below the velocity resolution limit, while only 21 $\pm$ 12\% (3/14) of disked binaries are below the limit.  This could indicate that among binaries with disks, the effects of disk-locking are reduced, though we must keep the small sample sizes in mind.  Among the diskless single and binary systems, 36 $\pm$ 6\% (36/101) of the WTTS single stars are below the resolution limit, compared to 28 $\pm$ 10\% (8/29) of the WTTS binaries.  Without the disk, we would expect tidal interactions to be the primary cause of higher rotation speeds for binaries in this comparison, but it is difficult to tell if there is a difference due to the small sample of WTTS binaries.  It may also be that 3 Myr is not enough time for the effects of any tidal interactions to become pronounced.

\subsection{Comparing Measured Radii to Stellar Evolution Model Predictions} \label{sec:model-comp}

We present the results of our statistical modeling technique to compare our geometric radii (radius estimated from rotation period and $v$~sin~$i$ measurements) to radii predicted by various solar-metallicity evolutionary models as a function of luminosity.  We consider $\rho$ = 1 to indicate that a model provides a good fit to the measured radii, while $\rho$ $>$ 1 means that on average, the model underestimates the radii (radius inflation) and $\rho$ $<$ 1 means the model overestimates the radii.  For this part of the analysis, we use the Model Comparison Sample.

\begin{figure}
    \centering
    \includegraphics[width=0.9\linewidth]{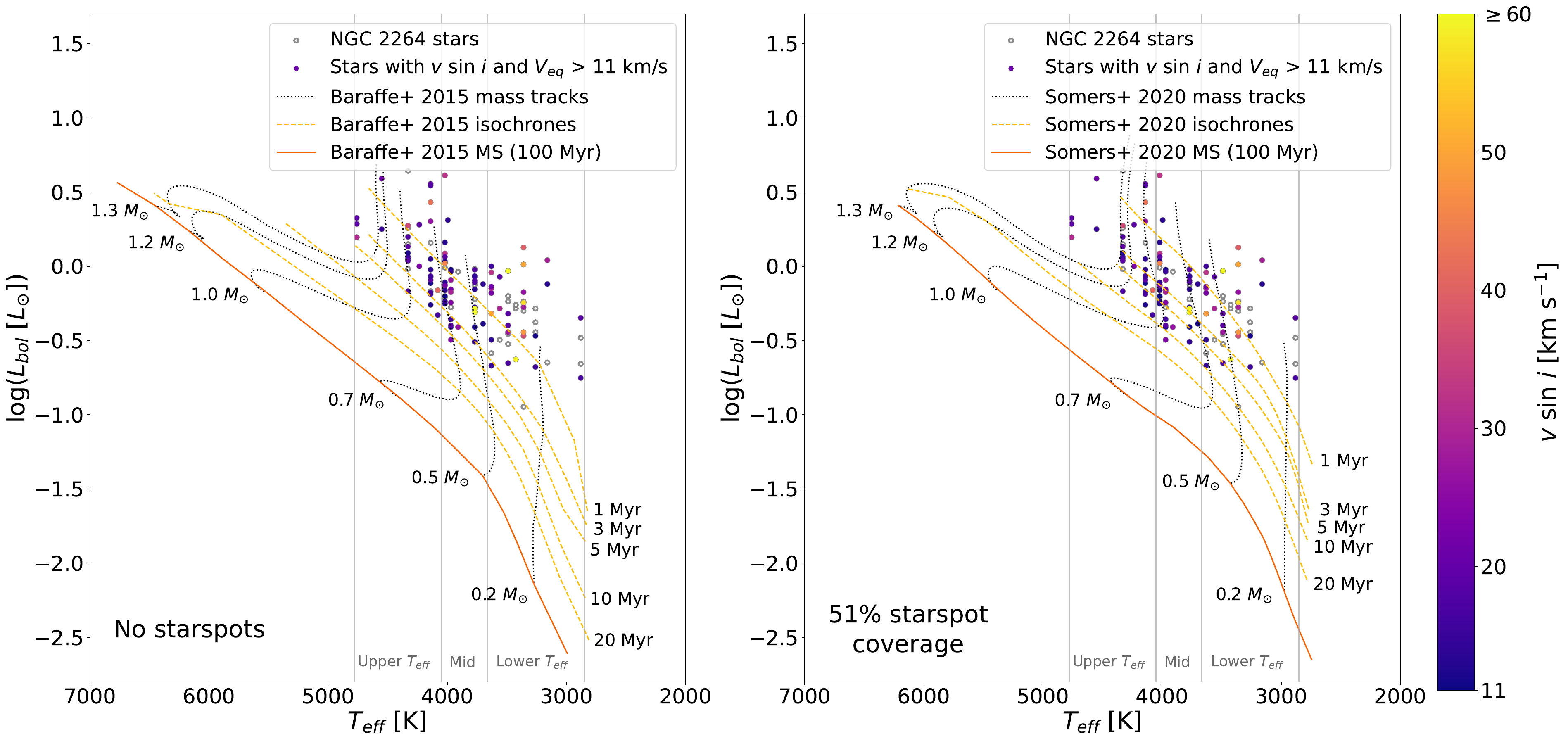}
    \caption{HR diagrams showing the Model Comparison Sample with isochrones (dashed yellow lines) and mass tracks (dotted black lines) from the starspot-free \cite{Baraffe2015} model (left) and the 51\% starspot coverage \cite{Somers2020} model (right).  The main sequence track for both models is approximated by the 100 Myr isochrone (solid orange line).  The color of the circles indicates the measured $v$~sin~$i$ of the stars, with stars with $v$~sin~$i$ or $V_{eq}$ below the 11 $\mathrm{km}\,\mathrm{s}^{-1}$ velocity resolution limit represented by open circles.  The vertical gray lines mark the main temperature-based subgroups that were also used in analysis.  The stars mostly appear well above the 3 Myr isochrone from the \cite{Baraffe2015} models, seeming closer to 1 Myr.  The stars are more scattered around the 3 Myr isochrone from the 51\% starspot coverage \cite{Somers2020} models, which is comparatively cooler than the \cite{Baraffe2015} 3 Myr isochrone.}
    \label{fig:HR_diagrams}
\end{figure}

We compare our radius measurements to predictions from two sets of stellar evolution models, \cite{Baraffe2015} and \cite{Somers2020}. Both models are solar metallicity, non-accreting, and appropriate for PMS stars.  The \cite{Baraffe2015} model consists of one set of starspot-free isochrones, which we employ for masses between 0.05 $M_{\odot}$ to 1.4 $M_{\odot}$.  The \cite{Somers2020} models have six sets of isochrones, corresponding to starspot covering fractions of 0\%, 17\%, 34\%, 51\%, 68\%, and 85\%, and with masses ranging from 0.1 $M_{\odot}$ to 1.3 $M_{\odot}$.  We compare the radius predictions for each set of models to the entire Model Comparison Sample (``all" objects), as well as to three temperature-based subgroups.  The ``all" group contains 136 objects, 100 of which have a measured value of $r$~sin~$i$ ($v$~sin~$i$ $>$ 11 $\mathrm{km}\,\mathrm{s}^{-1}$). The ``lower" group has 48 objects with $T_{\rm eff}$ ranging from 2880$-$3630 K, 29 of which have a measured value of $r$~sin~$i$.  The ``mid" group has 49 objects with $T_{\rm eff}$ ranging from 3631$-$4020 K, 40 of which have a measured value of $r$~sin~$i$.  The ``upper" group has 39 objects with $T_{\rm eff}$ ranging from 4021$-$4760 K, 31 of which have a measured value of $r$~sin~$i$.  Fig. \ref{fig:HR_diagrams} shows HR diagrams for the Model Comparison Sample, with isochrones and mass tracks for the \cite{Baraffe2015} model (left) and the 51\% starspot coverage \cite{Somers2020} model (right). The boundaries for the temperature subgroups (as a proxy for mass) are shown by the vertical gray lines.  In accordance with the estimated age of NGC 2264, we primarily used the 3 Myr isochrone for each set of models, though we also tested the 1 Myr and 5 Myr isochrones.

\subsubsection{Comparison to Baraffe et al. (2015) Model Radius Predictions}

We begin with the \cite{Baraffe2015} models, which assume the stars' photospheres are free of starspots.  Our results are listed in Table \ref{tab:Baraffe_rho}. For the 3 Myr model, we find that the objects have an average $\rho$ = 1.200 $\pm$ 0.020, indicating an over-radius much like \cite{Jackson2018} found for the 120 Myr Pleiades stars.  When we look at the temperature/mass sub-groups, we see that the ``lower" and ``mid" groups are roughly in agreement about how much the model is underestimating the radii and consistent with the mean $\rho$.  However, the ``upper" group appears to be well-estimated by the 3 Myr model, with $\rho$ = 1.001 $\pm$ 0.028. 

We conducted a few tests to explore how dependent these results were on the sub-group divisions.  First, we redefined the bins so that each temperature subgroup spanned an equal range of temperatures ($\sim$627 K).  The ``upper-eq" group remained largely unchanged, but several objects moved from the ``lower-eq" group to the ``mid-eq" group.  Despite this, the $\rho$ for the ``mid-eq" and ``lower-eq" groups also remained unchanged.  Even though grouping by luminosity is not as strongly correlated with mass as temperature is, we also tried splitting the objects into three luminosity-based groups, as was done in \cite{Jackson2018}.  The ``lower-lum" and ``mid-lum" groups showed an average over-radius consistent with the mean for the entire set of objects ($\rho$ = 1.21 $\pm$ 0.03), but the ``upper-lum" group had $\rho$ = 1.161 $\pm$ 0.037.  This is much closer to the total mean over-radius, but still not as over-inflated as the other two groups, suggesting that the largest and brightest objects are fairly well described by the \cite{Baraffe2015} 3 Myr model.  Finally, we split each of our original three temperature groups in half, to further explore a potential relationship with the mass and check whether extreme behavior on the outer edges of the temperature range was being disguised by the group averages.  Both halves of the ``mid" group were in agreement with each other and the ``mid" group average, as expected.  The same was true for both halves of the ``upper" group.  For the ``lower" group, both halves agreed with the ``lower" group average, but were not in agreement with each other, with the lowest group showing significant radius inflation of $\rho$ = 1.496 $\pm$ 0.267.  This suggests that there may be some mass dependence to the degree of radius inflation, at least for the very lowest and very highest masses in our sample. 

We then tested the 1 Myr \cite{Baraffe2015} model and got a mean $\rho$ of 1.006 $\pm$ 0.019, with the ``lower" and ``mid" groups being well-described by the model and the ``upper" group radii being ``under-inflated" compared to the model prediction ($\rho$ = 0.845 $\pm$ 0.016).  For the 5 Myr model, the mean $\rho$ was 1.305 $\pm$ 0.028, and all subgroups showed an over-radius; $\sim$25\% and $\sim$35\% for the ``lower" and ``mid" groups respectively, but only about 11\% for the ``upper" group.

\begin{deluxetable}{c|cc|cc|cc|cc|}
\tablecaption{Radius ratio, $\rho$, between radii measured for low-mass PMS stars in NGC 2264 and radii predicted by the \cite{Baraffe2015} evolutionary models at 1, 3, and 5 Myr. ln $\widehat{\mathcal{L}}$ is the maximum of the log-likelihood function.  The numbers below each temperature group indicate the total number of objects in that group, with the number of objects with a measured value of $r$~sin~$i$ in parentheses.}
\label{tab:Baraffe_rho}
\tablehead{\colhead{\textbf{Model Age}} & \multicolumn{2}{c}{\textbf{All}} & \multicolumn{2}{c}{\textbf{Lower}} & \multicolumn{2}{c}{\textbf{Mid}} & \multicolumn{2}{c}{\textbf{Upper}} \\ 
\colhead{} & \multicolumn{2}{c}{136 (100)} & \multicolumn{2}{c}{48 (29)} & \multicolumn{2}{c}{49 (40)} & \multicolumn{2}{c}{39 (31)} \\
\colhead{} & \colhead{ln $\widehat{\mathcal{L}}$} & \colhead{$\rho$} & \colhead{ln $\widehat{\mathcal{L}}$} & \colhead{$\rho$} & \colhead{ln $\widehat{\mathcal{L}}$} & \colhead{$\rho$} & \colhead{ln $\widehat{\mathcal{L}}$} & \colhead{$\rho$} }

\startdata
\textbf{1 Myr} & -43.973 & 1.006 $\pm$ 0.019 & -31.384 & 0.978 $\pm$ 0.045 & -6.135 & 1.040 $\pm$ 0.024 & 7.201 & 0.845 $\pm$ 0.016 \\
\hline
\textbf{3 Myr} & -71.371 & 1.200 $\pm$ 0.020 & -39.840 & 1.146 $\pm$ 0.080 & -16.310 & 1.241 $\pm$ 0.027 & -3.174 & 1.001 $\pm$ 0.028 \\
\hline
\textbf{5 Myr} & -79.570 & 1.305 $\pm$ 0.028 & -43.869 & 1.249 $\pm$ 0.088 & -18.945 & 1.346 $\pm$ 0.028 & -3.293 & 1.105 $\pm$ 0.016 \\
\enddata
\end{deluxetable}

\subsubsection{The Influence of Starspots and Comparison to Somers et al. (2020) Model Radius Predictions}

At this point, we define two separate but related problems.  The first is the difference in the degree of radius inflation for the upper and lower mass stars across all model ages.  This could be interpreted as a difference in the model's effectiveness at estimating the radius at different masses, or as a mass-dependent age difference.  It has been noted by \cite{Hartmann2003} that in young associations, age estimation has a mass dependence, where ages estimated for 1 to 2 $M_{\odot}$ stars tend to be higher than ages for lower mass stars in the same association.  This trend has been observed in Upper Scorpius, IC 348, and the Orion Nebula Complex, as well as NGC 2264 \citep{Palla2000}. \cite{Hartmann2003} suggests that this is due to differences in birthline effects at different masses, leading to overestimated ages for stars with temperatures exceeding 4000 K and/or luminosities brighter than 1 $L_{\odot}$, relative to derived ages for lower mass stars \citep{Hosokawa2011, Hartmann2016}.  The majority of the ``upper" group stars are in this region of the HR diagram, which could explain why it's better fit by the older model compared to the ``mid" and ``lower" groups. In their analysis of the Pleiades cluster, \cite{Jackson2018} did not identify a difference in the degree of radius inflation between different luminosity bins (which they used as a proxy for mass), but very few of their stars had temperatures above 4000 K, so any differences from birthline effects may have been averaged out of the highest luminosity bin. The Pleiades are also significantly older than the clusters listed in \cite{Hartmann2003}, so it may be that by 120 Myr, any differences from the birthlines at different masses have been significantly reduced.

The second issue is whether the radius inflation in the ``mid" and ``lower" groups is the result of using an incorrect age, or if the model temperatures are incorrect.  Our analysis method assumes a model radius and implicitly a $T_{\rm eff}$ based on the luminosity and isochrone used.  When that model radius is compared to the geometric radius (from the period and $v$ sin $i$), the model is statistically smaller, which means either the assumed $T_{\rm eff}$ is too hot or the isochrone is too old.  \cite{Kraus2015} compared the properties of the eclipsing binary system UScoCTIO 5 to predictions from \cite{Baraffe2015}.  They found that the temperature predicted by the model isomass line was too high for the estimated temperatures of the components at any age on the HR diagram.  They also found that at the age of the host cluster (estimated from high mass main sequence stars), the model reproduced an appropriate mass on the mass-radius diagram, but underestimated the radius; another indication that the temperatures from the model were too high \citep{Kraus2015}.  

\cite{Somers2015a} have proposed that including the effects of starspots in modeling can mitigate the issue of radius inflation.  In effect, a spotless model would overestimate the temperature of a spotted star at a given luminosity and age, which leads to underestimating the radius. Starspots primarily decrease the effective temperature of a star without significantly affecting the luminosity, which increases the predicted radius of the star \citep{Somers2020}.  T Tauri stars are known to be very magnetically active, which increases starspot activity.  Introducing starspots may also solve the mass-dependent age disagreement.  \cite{Somers2020} demonstrated that discrepant ages derived for low and higher mass groups of PMS stars in Upper Scorpius were in better agreement as the starspot fraction increased.  In addition to this, the average derived ages for both groups increased overall, indicating that spotless models had underestimated the ages of the PMS stars.

\begin{deluxetable}{cc|cc|cc|cc|cc|}
\tablecaption{Radius ratio, $\rho$, between radii measured for low-mass PMS stars in NGC 2264 and radii predicted by the \cite{Somers2020} evolutionary models at 1, 3, and 5 Myr.  ln $\widehat{\mathcal{L}}$ is the maximum of the log-likelihood function. The numbers below each temperature group indicate the total number of objects in that group, with the number of objects with a measured value of $r$~sin~$i$ in parentheses.}
\label{tab:SPOTS_rho}
\tablehead{\colhead{\textbf{Model Age}} & \colhead{\textbf{Starspot Fraction}} & \multicolumn{2}{c}{\textbf{All}} & \multicolumn{2}{c}{\textbf{Lower}} & \multicolumn{2}{c}{\textbf{Mid}} & \multicolumn{2}{c}{\textbf{Upper}} \\ 
\colhead{} & \colhead{} & \multicolumn{2}{c}{136 (100)} & \multicolumn{2}{c}{48 (29)} & \multicolumn{2}{c}{49 (40)} & \multicolumn{2}{c}{39 (31)} \\
\colhead{} & \colhead{} & \colhead{ln $\widehat{\mathcal{L}}$} & \colhead{$\rho$} & \colhead{ln $\widehat{\mathcal{L}}$} & \colhead{$\rho$} & \colhead{ln $\widehat{\mathcal{L}}$} & \colhead{$\rho$} & \colhead{ln $\widehat{\mathcal{L}}$} & \colhead{$\rho$} }

\startdata
\textbf{1 Myr} & 0\% & -42.063 & 0.986 $\pm$ 0.019 & -30.025 & 0.967 $\pm$ 0.046 & -5.872 & 1.018 $\pm$ 0.021 & 7.505 & 0.835 $\pm$ 0.016 \\
 & 17\% & -37.165 & 0.954 $\pm$ 0.018 & -28.584 & 0.948 $\pm$ 0.050 & -3.993 & 0.984 $\pm$ 0.020 & 8.816 & 0.804 $\pm$ 0.016 \\
 & 34\% & -32.490 & 0.919 $\pm$ 0.018 & -27.086 & 0.882 $\pm$ 0.070 & -2.377 & 0.949 $\pm$ 0.021 & 9.293 & 0.772 $\pm$ 0.022 \\
 & 51\% & -26.957 & 0.888 $\pm$ 0.015 & -25.089 & 0.843 $\pm$ 0.063 & -0.282 & 0.912 $\pm$ 0.019 & 12.420 & 0.743 $\pm$ 0.007 \\
 & 68\% & -21.618 & 0.849 $\pm$ 0.018 & -23.009 & 0.802 $\pm$ 0.061 & 1.609 & 0.874 $\pm$ 0.019 & 13.670 & 0.702 $\pm$ 0.012 \\
 & 85\% & -15.513 & 0.818 $\pm$ 0.016 & -20.999 & 0.776 $\pm$ 0.055 & 3.907 & 0.843 $\pm$ 0.018 & 15.396 & 0.674 $\pm$ 0.009 \\
\hline
\textbf{3 Myr} & 0\% & -69.045 & 1.209 $\pm$ 0.024 & -40.555 & 1.150 $\pm$ 0.079 & -15.001 & 1.250 $\pm$ 0.026 & -2.003 & 1.043 $\pm$ 0.025 \\
 & 17\% & -62.952 & 1.171 $\pm$ 0.021 & -38.735 & 1.110 $\pm$ 0.076 & -12.686 & 1.205 $\pm$ 0.025 & -5.757 & 1.066 $\pm$ 0.038 \\
 & 34\% & -58.912 & 1.123 $\pm$ 0.023 & -36.970 & 1.068 $\pm$ 0.073 & -11.338 & 1.159 $\pm$ 0.026 & 1.489 & 0.949 $\pm$ 0.021 \\
 & 51\% & -52.632 & 1.082 $\pm$ 0.022 & -35.022 & 1.026 $\pm$ 0.065 & -8.950 & 1.119 $\pm$ 0.025 & -2.912 & 0.985 $\pm$ 0.035 \\
 & 68\% & -49.508 & 1.039 $\pm$ 0.019 & -33.041 & 0.992 $\pm$ 0.053 & -8.171 & 1.077 $\pm$ 0.024 & 3.904 & 0.856 $\pm$ 0.015 \\
 & 85\% & -41.904 & 0.996 $\pm$ 0.020 & -31.081 & 0.973 $\pm$ 0.044 & -5.255 & 1.030 $\pm$ 0.023 & 7.206 & 0.848 $\pm$ 0.014 \\
\hline
\textbf{5 Myr} & 0\% & -79.106 & 1.302 $\pm$ 0.026 & -44.102 & 1.243 $\pm$ 0.083 & -18.605 & 1.347 $\pm$ 0.027 & -6.564 & 1.138 $\pm$ 0.043 \\
 & 17\% & -73.930 & 1.259 $\pm$ 0.024 & -42.433 & 1.200 $\pm$ 0.082 & -16.505 & 1.298 $\pm$ 0.028 & -9.975 & 1.093 $\pm$ 0.036 \\
 & 34\% & -70.074 & 1.212 $\pm$ 0.025 & -40.615 & 1.149 $\pm$ 0.079 & -15.365 & 1.256 $\pm$ 0.028 & -0.963 & 1.024 $\pm$ 0.023 \\
 & 51\% & -63.877 & 1.169 $\pm$ 0.023 & -38.852 & 1.112 $\pm$ 0.076 & -12.812 & 1.211 $\pm$ 0.026 & -7.911 & 1.070 $\pm$ 0.056 \\
 & 68\% & -59.683 & 1.122 $\pm$ 0.019 & -36.894 & 1.071 $\pm$ 0.072 & -11.951 & 1.163 $\pm$ 0.026 & 1.137 & 0.940 $\pm$ 0.027 \\
 & 85\% & -53.068 & 1.074 $\pm$ 0.022 & -34.89 & 1.023 $\pm$ 0.066 & -9.113 & 1.112 $\pm$ 0.025 & 4.551 & 0.906 $\pm$ 0.011 \\
\enddata

\end{deluxetable}

\begin{figure}
    \centering
    \includegraphics[width=0.9\linewidth]{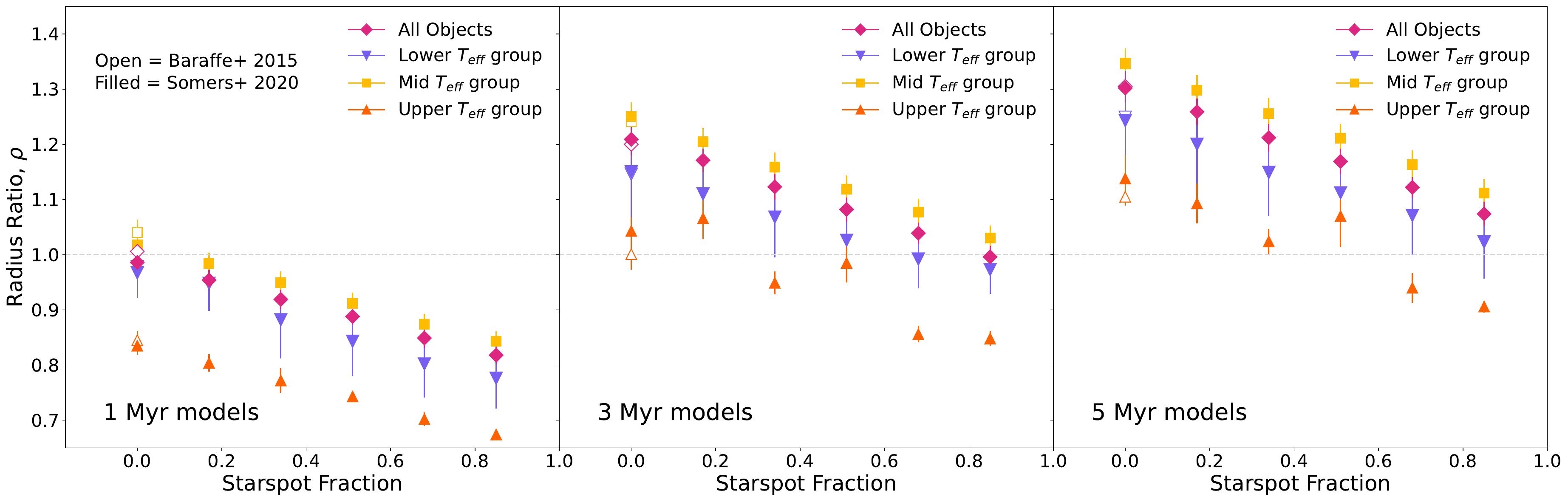}
    \caption{Radius ratio of radius from $v$~sin~$i$ and period measurements to radius predictions from models with different starspot coverage fractions, for 1 Myr (left), 3 Myr (center), and 5 Myr (right) models.  Results are shown by temperature subgroup; all objects (pink diamonds), lower $T_{\rm eff}$ group (purple downward-pointing triangles), mid $T_{\rm eff}$ group (yellow squares), and upper $T_{\rm eff}$ group (orange upward-pointing triangles).  The \cite{Baraffe2015} model is shown with open symbols (only for 0\% starspot fraction, as those models did not incorporate starspots) and the \cite{Somers2020} models use filled symbols.  For all groups, radius inflation decreases as starspot fraction increases.}
    \label{fig:rsini_rho}
\end{figure}

We tested predictions from all six starspot fraction models from \cite{Somers2020} against our measurements.  The results are listed in Table \ref{tab:SPOTS_rho}, and represented graphically in Fig. \ref{fig:rsini_rho}.  Some trends were present in the models at all three ages (1, 3, and 5 Myr).  The 0\% model had nearly-perfect agreement with the results from the \cite{Baraffe2015} model, which implies that any differences will be mainly a result of the starspots and not other differences in underlying assumptions between the models.  In general, as the starspot fraction increased, the average over-radius decreased for all groups.  For the ``lower" and ``mid" groups, this decrease was fairly linear at all three ages, while for the ``upper" group, it was only linear for the 1 Myr models.  For the 3 Myr models, the decreases in $\rho$ for the ``upper" group occurred in pairs; the 0\% and 17\% starspot fraction models had the same $\rho$, as did the 34\% and 51\% models, and the 68\% and 85\% models. For the 5 Myr models, $\rho$ followed a roughly linear trend except for at 51\%.

For the ``lower" and ``mid" groups, the radius inflation at 3 Myr appears to be significantly reduced for starspot fractions $>$ 68\%.  While this coverage is high, there has been other evidence that young, low-mass stars can be significantly covered in spots.  For example, in an investigation into the unusually blue colors of K dwarfs in the Pleiades, \cite{Stauffer2003} suggested that high stellar rotation may drive starspot formation, and found that the best fitting models for the starspot coverage had fractions in excess of 50\%.  More recently, \cite{Paolino2024} created composite models of ten WTTS stars in the Taurus$-$Auriga complex and determined that the appropriate starspot coverage ranged from 50\% to 85\% for the majority of the stars.  Their results are complimented by similar findings from \cite{Gangi2022}, who demonstrated that up to 90\% of the total flux for a selection of CTTS stars in the Taurus$-$Auriga complex came from cool regions on the stars.  It may also be true that NGC 2264 is slightly younger than 3 Myr, so using a younger isochrone would not necessitate such an extreme starspot coverage for the majority of stars in the cluster.  However, given the results of \cite{Paolino2024} and \cite{Gangi2022}, we would still expect the age to be closer to 3 Myr than 1 Myr, with significant starspot coverage.  

However, even with 85\% starspot coverage, we still see a difference in the degree of radius inflation between the ``upper" and ``lower" groups at all ages.  It may be that the younger the cluster is, the more significant the differences from birthline effects appear to be.  \cite{Hosokawa2011} compared HR diagram positions for stars modeled with various accretion histories prior to arriving on the pre-main sequence evolutionary tracks to isochrones from \cite{Baraffe1998}.  They found that ages for higher mass stars were more overestimated than ages from lower mass stars at 0.3 Myr, no matter the accretion history.  As the stars aged, the relative age gap between lower and higher mass stars decreased, though was still present at 10 Myr \citep{Hosokawa2011}.  Upper Scorpius is older than NGC 2264; \cite{Somers2020} started with an estimate of 5 Myr and converged around 6$-$7 Myr with the inclusion of starspots, but mass and radius measurements of eclipsing binary members by \cite{Kraus2015} and \cite{David2019} suggest that the \cite{Pecaut2012} estimate of 10$-$11 Myr may be more appropriate.  It is possible that by this age, the differences from birthline effects have reduced enough that the inclusion of starspots closes the gap.

\section{Summary and Conclusions} \label{sec:conclusion}

Using Fourier cross-correlation, we have measured the projected rotational velocity, $v$~sin~$i$, for 254 low-mass, PMS stars that are members of the $\sim$3 Myr old open cluster NGC 2264.  These measurements provide lower limits on rotation velocity and, when combined with rotation period measurements, lower limits on radius for these stars.  As NGC 2264 is a well-studied cluster, we have taken advantage of years of auxiliary data to help identify properties of the stellar systems such as circumstellar disk status and binarity.  We explored the $v$~sin~$i$ distributions for stars grouped by these properties to see if they may have an effect on the angular momentum of the stars.  We have also used a maximum likelihood method to estimate the average radius of stars in the cluster compared to radius predictions from two sets of stellar evolution models; one starspot-free \citep{Baraffe2015}, the other incorporating a range of possible starspot proportions on the stellar surface \citep{Somers2020}. The main findings of this work are summarized as follows:
\begin{enumerate}
    \item We estimate that about 38\% of low-mass stars in NGC 2264 show evidence of a circumstellar disk.  Our analysis was limited by the velocity resolution limit of our data, but it indicates that stars with disks may rotate slower than diskless stars.  The most likely mechanism proposed for angular momentum loss caused by interaction between the star and the disk is disk-locking through magnetic field lines (e.g., \citealt{Konigl1991, Cieza2007b, Bouvier2014}).
    \item Accurate measurement of $v$~sin~$i$ in binary systems is complicated by several factors, which can result in over-estimated $v$~sin~$i$ values, but our results suggest that stars in close binary systems may rotate faster than single stars.  It is possible that stars in binary systems may be more likely to have disrupted circumstellar disks, limiting the potential effects of disk-locking, or that tidal interactions between the close stars could spin them up, though our small sample sizes and the significant youth of NGC 2264 make it difficult for us to explore these options.
    \item After comparing geometrical radii derived from the period and $v$~sin~$i$ to the radii predicted by stellar evolutionary models with statistical analysis methods, we report that the stellar radii of our sample stars in NGC 2264 appear to be larger than the 3 Myr model predictions from \cite{Baraffe2015}, by about 20 $\pm$ 2\%.  Using temperature as a proxy for mass, we also observe a difference in the degree of radius inflation between the lower mass stars and higher mass stars, with the lower mass stars being more inflated.  It is possible that this is a result of differences in the birthline for stars of different masses, and that these effects will become reduced as the cluster ages.  We demonstrate that models which incorporate the effects of starspots, such as those of \cite{Somers2020}, are more successful at reproducing the observed radii at the age of 3 Myr for the lower mass stars.  There is some evidence to suggest that a large fraction ($>$50\%) of the stellar surface in these low-mass, magnetically active stars is covered in starspots.
\end{enumerate}

The first 15 Myr in the pre-main sequence phase are a time of intense activity for a low-mass star: the luminosity, temperature, and radius change rapidly as the star contracts, magnetic fields drive accretion and jets, and circumstellar disks dissipate.  Beginning with NGC 2264, we are conducting a detailed study of rotational evolution in T Tauri stars through this stage in their development.  We have selected an additional four clusters with ages ranging from 1 to 14 Myr on which to perform similar analyses as we have done in this work.  We have also illustrated the need for high-resolution spectroscopy to discern the influences that circumstellar disks and binarity may have on the natural progression of angular momentum in the early stages of a star's lifetime.  Additionally, when combined with rotation periods, accurate measurements of projected rotation velocity can be used to estimate the physical properties of a cluster of stars.  In the era of time-domain photometric surveys such as the Transiting Exoplanet Survey Satellite (TESS) and Zwicky Transient Facility (ZTF) contributing tens of thousands of new rotation period measurements, complementary rotation velocity measurements can be a powerful tool to evaluate the effectiveness of stellar evolutionary models.  

\begin{acknowledgments}
The authors would like to thank the anonymous referee for their thoughtful comments, which improved the quality of the manuscript.

LMG acknowledges support from the Indiana University (IU) Astronomy Department Sullivan Fellowship and the IU College of Arts and Sciences Dissertation Research Fellowship.  We thank the Indiana University College of Arts and Sciences for funding IU's share of the WIYN telescope.  We thank the staff of the WIYN Observatory and Kitt Peak National Observatory for their help and support during our WIYN Hydra observing runs.

This work has made use of data from the European Space Agency (ESA) mission
{\it Gaia} (\url{https://www.cosmos.esa.int/gaia}), processed by the {\it Gaia}
Data Processing and Analysis Consortium (DPAC,
\url{https://www.cosmos.esa.int/web/gaia/dpac/consortium}). Funding for the DPAC
has been provided by national institutions, in particular the institutions
participating in the {\it Gaia} Multilateral Agreement.

This research made use of the open source Python package exoctk, the Exoplanet Characterization Toolkit \citep{Bourque2021}.

\end{acknowledgments}

\vspace{5mm}
\facilities{WIYN(Hydra)}

\software{IRAF \citep{Tody1986, Tody1993}}


\bibliography{n2264_bib}{}

\begin{thebibliography}{}
\expandafter\ifx\csname natexlab\endcsname\relax\def\natexlab#1{#1}\fi
\providecommand{\url}[1]{\href{#1}{#1}}
\providecommand{\dodoi}[1]{doi:~\href{http://doi.org/#1}{\nolinkurl{#1}}}
\providecommand{\doeprint}[1]{\href{http://ascl.net/#1}{\nolinkurl{http://ascl.net/#1}}}
\providecommand{\doarXiv}[1]{\href{https://arxiv.org/abs/#1}{\nolinkurl{https://arxiv.org/abs/#1}}}

\bibitem[{{Affer} {et~al.}(2013){Affer}, {Micela}, {Favata}, {Flaccomio}, \& {Bouvier}}]{Affer2013}
{Affer}, L., {Micela}, G., {Favata}, F., {Flaccomio}, E., \& {Bouvier}, J. 2013, \mnras, 430, 1433, \dodoi{10.1093/mnras/stt003}

\bibitem[{{Aizawa} {et~al.}(2020){Aizawa}, {Suto}, {Oya}, {Ikeda}, \& {Nakazato}}]{Aizawa2020}
{Aizawa}, M., {Suto}, Y., {Oya}, Y., {Ikeda}, S., \& {Nakazato}, T. 2020, \apj, 899, 55, \dodoi{10.3847/1538-4357/aba43d}

\bibitem[{{Bachiller}(1996)}]{Bachiller1996}
{Bachiller}, R. 1996, \araa, 34, 111, \dodoi{10.1146/annurev.astro.34.1.111}

\bibitem[{{Baraffe} {et~al.}(1998){Baraffe}, {Chabrier}, {Allard}, \& {Hauschildt}}]{Baraffe1998}
{Baraffe}, I., {Chabrier}, G., {Allard}, F., \& {Hauschildt}, P.~H. 1998, \aap, 337, 403, \dodoi{10.48550/arXiv.astro-ph/9805009}

\bibitem[{{Baraffe} {et~al.}(2015){Baraffe}, {Homeier}, {Allard}, \& {Chabrier}}]{Baraffe2015}
{Baraffe}, I., {Homeier}, D., {Allard}, F., \& {Chabrier}, G. 2015, \aap, 577, A42, \dodoi{10.1051/0004-6361/201425481}

\bibitem[{{Barnes}(2003)}]{Barnes2003}
{Barnes}, S.~A. 2003, \apj, 586, 464, \dodoi{10.1086/367639}

\bibitem[{{Bastian} {et~al.}(2020){Bastian}, {Kamann}, {Amard}, {Charbonnel}, {Haemmerl{\'e}}, \& {Matt}}]{Bastian2020}
{Bastian}, N., {Kamann}, S., {Amard}, L., {et~al.} 2020, \mnras, 495, 1978, \dodoi{10.1093/mnras/staa1332}

\bibitem[{{Baxter} {et~al.}(2009){Baxter}, {Covey}, {Muench}, {F{\H{u}}r{\'e}sz}, {Rebull}, \& {Szentgyorgyi}}]{Baxter2009}
{Baxter}, E.~J., {Covey}, K.~R., {Muench}, A.~A., {et~al.} 2009, \aj, 138, 963, \dodoi{10.1088/0004-6256/138/3/963}

\bibitem[{{Bertout}(1989)}]{Bertout1989}
{Bertout}, C. 1989, \araa, 27, 351, \dodoi{10.1146/annurev.aa.27.090189.002031}

\bibitem[{Bourque {et~al.}(2021)Bourque, Espinoza, Filippazzo, Fix, King, Martlin, Medina, Batalha, Fox, Fowler, Fraine, Hill, Lewis, Stevenson, Valenti, \& Wakeford}]{Bourque2021}
Bourque, M., Espinoza, N., Filippazzo, J., {et~al.} 2021, The Exoplanet Characterization Toolkit (ExoCTK), 1.0.0,  Zenodo, \dodoi{10.5281/zenodo.4556063}

\bibitem[{{Bouvier}(1991)}]{Bouvier1991}
{Bouvier}, J. 1991, in NATO Advanced Study Institute (ASI) Series C, Vol. 340, Angular Momentum Evolution of Young Stars, ed. S.~{Catalano} \& J.~R. {Stauffer}, 41

\bibitem[{{Bouvier} {et~al.}(1993){Bouvier}, {Cabrit}, {Fernandez}, {Martin}, \& {Matthews}}]{Bouvier1993}
{Bouvier}, J., {Cabrit}, S., {Fernandez}, M., {Martin}, E.~L., \& {Matthews}, J.~M. 1993, \aap, 272, 176

\bibitem[{{Bouvier} {et~al.}(2014){Bouvier}, {Matt}, {Mohanty}, {Scholz}, {Stassun}, \& {Zanni}}]{Bouvier2014}
{Bouvier}, J., {Matt}, S.~P., {Mohanty}, S., {et~al.} 2014, in Protostars and Planets VI, ed. H.~{Beuther}, R.~S. {Klessen}, C.~P. {Dullemond}, \& T.~{Henning}, 433--450, \dodoi{10.2458/azu_uapress_9780816531240-ch019}

\bibitem[{{Brice{\~n}o} {et~al.}(2019){Brice{\~n}o}, {Calvet}, {Hern{\'a}ndez}, {Vivas}, {Mateu}, {Downes}, {Loerincs}, {P{\'e}rez-Blanco}, {Berlind}, {Espaillat}, {Allen}, {Hartmann}, {Mateo}, \& {Bailey}}]{Briceno2019}
{Brice{\~n}o}, C., {Calvet}, N., {Hern{\'a}ndez}, J., {et~al.} 2019, \aj, 157, 85, \dodoi{10.3847/1538-3881/aaf79b}

\bibitem[{{Chandrasekhar} \& {M{\"u}nch}(1950)}]{Chandrasekhar1950}
{Chandrasekhar}, S., \& {M{\"u}nch}, G. 1950, \apj, 111, 142, \dodoi{10.1086/145245}

\bibitem[{{Cieza} \& {Baliber}(2006)}]{Cieza2006}
{Cieza}, L., \& {Baliber}, N. 2006, \apj, 649, 862, \dodoi{10.1086/506342}

\bibitem[{{Cieza} \& {Baliber}(2007)}]{Cieza2007b}
---. 2007, \apj, 671, 605, \dodoi{10.1086/522080}

\bibitem[{{Cieza} {et~al.}(2007){Cieza}, {Padgett}, {Stapelfeldt}, {Augereau}, {Harvey}, {Evans}, {Mer{\'\i}n}, {Koerner}, {Sargent}, {van Dishoeck}, {Allen}, {Blake}, {Brooke}, {Chapman}, {Huard}, {Lai}, {Mundy}, {Myers}, {Spiesman}, \& {Wahhaj}}]{Cieza2007a}
{Cieza}, L., {Padgett}, D.~L., {Stapelfeldt}, K.~R., {et~al.} 2007, \apj, 667, 308, \dodoi{10.1086/520698}

\bibitem[{{Clarke} \& {Bouvier}(2000)}]{Clarke2000}
{Clarke}, C.~J., \& {Bouvier}, J. 2000, \mnras, 319, 457, \dodoi{10.1046/j.1365-8711.2000.03855.x}

\bibitem[{{Cody} {et~al.}(2014){Cody}, {Stauffer}, {Baglin}, {Micela}, {Rebull}, {Flaccomio}, {Morales-Calder{\'o}n}, {Aigrain}, {Bouvier}, {Hillenbrand}, {Gutermuth}, {Song}, {Turner}, {Alencar}, {Zwintz}, {Plavchan}, {Carpenter}, {Findeisen}, {Carey}, {Terebey}, {Hartmann}, {Calvet}, {Teixeira}, {Vrba}, {Wolk}, {Covey}, {Poppenhaeger}, {G{\"u}nther}, {Forbrich}, {Whitney}, {Affer}, {Herbst}, {Hora}, {Barrado}, {Holtzman}, {Marchis}, {Wood}, {Medeiros Guimar{\~a}es}, {Lillo Box}, {Gillen}, {McQuillan}, {Espaillat}, {Allen}, {D'Alessio}, \& {Favata}}]{Cody2014}
{Cody}, A.~M., {Stauffer}, J., {Baglin}, A., {et~al.} 2014, \aj, 147, 82, \dodoi{10.1088/0004-6256/147/4/82}

\bibitem[{{Corsaro} {et~al.}(2017){Corsaro}, {Lee}, {Garc{\'\i}a}, {Hennebelle}, {Mathur}, {Beck}, {Mathis}, {Stello}, \& {Bouvier}}]{Corsaro2017}
{Corsaro}, E., {Lee}, Y.-N., {Garc{\'\i}a}, R.~A., {et~al.} 2017, Nature Astronomy, 1, 0064, \dodoi{10.1038/s41550-017-0064}

\bibitem[{{David} {et~al.}(2019){David}, {Hillenbrand}, {Gillen}, {Cody}, {Howell}, {Isaacson}, \& {Livingston}}]{David2019}
{David}, T.~J., {Hillenbrand}, L.~A., {Gillen}, E., {et~al.} 2019, \apj, 872, 161, \dodoi{10.3847/1538-4357/aafe09}

\bibitem[{{Dias} {et~al.}(2021){Dias}, {Monteiro}, {Moitinho}, {L{\'e}pine}, {Carraro}, {Paunzen}, {Alessi}, \& {Villela}}]{Dias2021}
{Dias}, W.~S., {Monteiro}, H., {Moitinho}, A., {et~al.} 2021, \mnras, 504, 356, \dodoi{10.1093/mnras/stab770}

\bibitem[{{Edwards} {et~al.}(1994){Edwards}, {Hartigan}, {Ghandour}, \& {Andrulis}}]{Edwards1994}
{Edwards}, S., {Hartigan}, P., {Ghandour}, L., \& {Andrulis}, C. 1994, \aj, 108, 1056, \dodoi{10.1086/117134}

\bibitem[{{Edwards} {et~al.}(1993){Edwards}, {Strom}, {Hartigan}, {Strom}, {Hillenbrand}, {Herbst}, {Attridge}, {Merrill}, {Probst}, \& {Gatley}}]{Edwards1993}
{Edwards}, S., {Strom}, S.~E., {Hartigan}, P., {et~al.} 1993, \aj, 106, 372, \dodoi{10.1086/116646}

\bibitem[{{Ekstr{\"o}m} {et~al.}(2012){Ekstr{\"o}m}, {Georgy}, {Eggenberger}, {Meynet}, {Mowlavi}, {Wyttenbach}, {Granada}, {Decressin}, {Hirschi}, {Frischknecht}, {Charbonnel}, \& {Maeder}}]{Ekstrom2012}
{Ekstr{\"o}m}, S., {Georgy}, C., {Eggenberger}, P., {et~al.} 2012, \aap, 537, A146, \dodoi{10.1051/0004-6361/201117751}

\bibitem[{{Fedele} {et~al.}(2010){Fedele}, {van den Ancker}, {Henning}, {Jayawardhana}, \& {Oliveira}}]{Fedele2010}
{Fedele}, D., {van den Ancker}, M.~E., {Henning}, T., {Jayawardhana}, R., \& {Oliveira}, J.~M. 2010, \aap, 510, A72, \dodoi{10.1051/0004-6361/200912810}

\bibitem[{{Feigelson} \& {Montmerle}(1999)}]{Feigelson1999}
{Feigelson}, E.~D., \& {Montmerle}, T. 1999, \araa, 37, 363, \dodoi{10.1146/annurev.astro.37.1.363}

\bibitem[{{F{\H{u}}r{\'e}sz} {et~al.}(2006){F{\H{u}}r{\'e}sz}, {Hartmann}, {Szentgyorgyi}, {Ridge}, {Rebull}, {Stauffer}, {Latham}, {Conroy}, {Fabricant}, \& {Roll}}]{Furesz2006}
{F{\H{u}}r{\'e}sz}, G., {Hartmann}, L.~W., {Szentgyorgyi}, A.~H., {et~al.} 2006, \apj, 648, 1090, \dodoi{10.1086/506140}

\bibitem[{{Flaccomio} {et~al.}(2000){Flaccomio}, {Micela}, {Sciortino}, {Damiani}, {Favata}, {Harnden}, \& {Schachter}}]{Flaccomio2000}
{Flaccomio}, E., {Micela}, G., {Sciortino}, S., {et~al.} 2000, \aap, 355, 651

\bibitem[{{Fleming} {et~al.}(2019){Fleming}, {Barnes}, {Davenport}, \& {Luger}}]{Fleming2019}
{Fleming}, D.~P., {Barnes}, R., {Davenport}, J. R.~A., \& {Luger}, R. 2019, \apj, 881, 88, \dodoi{10.3847/1538-4357/ab2ed2}

\bibitem[{{Gaia Collaboration} {et~al.}(2021){Gaia Collaboration}, {Brown}, {Vallenari}, {Prusti}, {de Bruijne}, {Babusiaux}, {Biermann}, {Creevey}, {Evans}, {Eyer}, {Hutton}, {Jansen}, {Jordi}, {Klioner}, {Lammers}, {Lindegren}, {Luri}, {Mignard}, {Panem}, {Pourbaix}, {Randich}, {Sartoretti}, {Soubiran}, {Walton}, {Arenou}, {Bailer-Jones}, {Bastian}, {Cropper}, {Drimmel}, {Katz}, {Lattanzi}, {van Leeuwen}, {Bakker}, {Cacciari}, {Casta{\~n}eda}, {De Angeli}, {Ducourant}, {Fabricius}, {Fouesneau}, {Fr{\'e}mat}, {Guerra}, {Guerrier}, {Guiraud}, {Jean-Antoine Piccolo}, {Masana}, {Messineo}, {Mowlavi}, {Nicolas}, {Nienartowicz}, {Pailler}, {Panuzzo}, {Riclet}, {Roux}, {Seabroke}, {Sordo}, {Tanga}, {Th{\'e}venin}, {Gracia-Abril}, {Portell}, {Teyssier}, {Altmann}, {Andrae}, {Bellas-Velidis}, {Benson}, {Berthier}, {Blomme}, {Brugaletta}, {Burgess}, {Busso}, {Carry}, {Cellino}, {Cheek}, {Clementini}, {Damerdji}, {Davidson}, {Delchambre}, {Dell'Oro}, {Fern{\'a}ndez-Hern{\'a}ndez}, {Galluccio}, {Garc{\'\i}a-Lario},
  {Garcia-Reinaldos}, {Gonz{\'a}lez-N{\'u}{\~n}ez}, {Gosset}, {Haigron}, {Halbwachs}, {Hambly}, {Harrison}, {Hatzidimitriou}, {Heiter}, {Hern{\'a}ndez}, {Hestroffer}, {Hodgkin}, {Holl}, {Jan{\ss}en}, {Jevardat de Fombelle}, {Jordan}, {Krone-Martins}, {Lanzafame}, {L{\"o}ffler}, {Lorca}, {Manteiga}, {Marchal}, {Marrese}, {Moitinho}, {Mora}, {Muinonen}, {Osborne}, {Pancino}, {Pauwels}, {Petit}, {Recio-Blanco}, {Richards}, {Riello}, {Rimoldini}, {Robin}, {Roegiers}, {Rybizki}, {Sarro}, {Siopis}, {Smith}, {Sozzetti}, {Ulla}, {Utrilla}, {van Leeuwen}, {van Reeven}, {Abbas}, {Abreu Aramburu}, {Accart}, {Aerts}, {Aguado}, {Ajaj}, {Altavilla}, {{\'A}lvarez}, {{\'A}lvarez Cid-Fuentes}, {Alves}, {Anderson}, {Anglada Varela}, {Antoja}, {Audard}, {Baines}, {Baker}, {Balaguer-N{\'u}{\~n}ez}, {Balbinot}, {Balog}, {Barache}, {Barbato}, {Barros}, {Barstow}, {Bartolom{\'e}}, {Bassilana}, {Bauchet}, {Baudesson-Stella}, {Becciani}, {Bellazzini}, {Bernet}, {Bertone}, {Bianchi}, {Blanco-Cuaresma}, {Boch}, {Bombrun}, {Bossini},
  {Bouquillon}, {Bragaglia}, {Bramante}, {Breedt}, {Bressan}, {Brouillet}, {Bucciarelli}, {Burlacu}, {Busonero}, {Butkevich}, {Buzzi}, {Caffau}, {Cancelliere}, {C{\'a}novas}, {Cantat-Gaudin}, {Carballo}, {Carlucci}, {Carnerero}, {Carrasco}, {Casamiquela}, {Castellani}, {Castro-Ginard}, {Castro Sampol}, {Chaoul}, {Charlot}, {Chemin}, {Chiavassa}, {Cioni}, {Comoretto}, {Cooper}, {Cornez}, {Cowell}, {Crifo}, {Crosta}, {Crowley}, {Dafonte}, {Dapergolas}, {David}, {David}, {de Laverny}, {De Luise}, {De March}, {De Ridder}, {de Souza}, {de Teodoro}, {de Torres}, {del Peloso}, {del Pozo}, {Delbo}, {Delgado}, {Delgado}, {Delisle}, {Di Matteo}, {Diakite}, {Diener}, {Distefano}, {Dolding}, {Eappachen}, {Edvardsson}, {Enke}, {Esquej}, {Fabre}, {Fabrizio}, {Faigler}, {Fedorets}, {Fernique}, {Fienga}, {Figueras}, {Fouron}, {Fragkoudi}, {Fraile}, {Franke}, {Gai}, {Garabato}, {Garcia-Gutierrez}, {Garc{\'\i}a-Torres}, {Garofalo}, {Gavras}, {Gerlach}, {Geyer}, {Giacobbe}, {Gilmore}, {Girona}, {Giuffrida}, {Gomel}, {Gomez},
  {Gonzalez-Santamaria}, {Gonz{\'a}lez-Vidal}, {Granvik}, {Guti{\'e}rrez-S{\'a}nchez}, {Guy}, {Hauser}, {Haywood}, {Helmi}, {Hidalgo}, {Hilger}, {H{\l}adczuk}, {Hobbs}, {Holland}, {Huckle}, {Jasniewicz}, {Jonker}, {Juaristi Campillo}, {Julbe}, {Karbevska}, {Kervella}, {Khanna}, {Kochoska}, {Kontizas}, {Kordopatis}, {Korn}, {Kostrzewa-Rutkowska}, {Kruszy{\'n}ska}, {Lambert}, {Lanza}, {Lasne}, {Le Campion}, {Le Fustec}, {Lebreton}, {Lebzelter}, {Leccia}, {Leclerc}, {Lecoeur-Taibi}, {Liao}, {Licata}, {Lindstr{\o}m}, {Lister}, {Livanou}, {Lobel}, {Madrero Pardo}, {Managau}, {Mann}, {Marchant}, {Marconi}, {Marcos Santos}, {Marinoni}, {Marocco}, {Marshall}, {Martin Polo}, {Mart{\'\i}n-Fleitas}, {Masip}, {Massari}, {Mastrobuono-Battisti}, {Mazeh}, {McMillan}, {Messina}, {Michalik}, {Millar}, {Mints}, {Molina}, {Molinaro}, {Moln{\'a}r}, {Montegriffo}, {Mor}, {Morbidelli}, {Morel}, {Morris}, {Mulone}, {Munoz}, {Muraveva}, {Murphy}, {Musella}, {Noval}, {Ord{\'e}novic}, {Orr{\`u}}, {Osinde}, {Pagani}, {Pagano},
  {Palaversa}, {Palicio}, {Panahi}, {Pawlak}, {Pe{\~n}alosa Esteller}, {Penttil{\"a}}, {Piersimoni}, {Pineau}, {Plachy}, {Plum}, {Poggio}, {Poretti}, {Poujoulet}, {Pr{\v{s}}a}, {Pulone}, {Racero}, {Ragaini}, {Rainer}, {Raiteri}, {Rambaux}, {Ramos}, {Ramos-Lerate}, {Re Fiorentin}, {Regibo}, {Reyl{\'e}}, {Ripepi}, {Riva}, {Rixon}, {Robichon}, {Robin}, {Roelens}, {Rohrbasser}, {Romero-G{\'o}mez}, {Rowell}, {Royer}, {Rybicki}, {Sadowski}, {Sagrist{\`a} Sell{\'e}s}, {Sahlmann}, {Salgado}, {Salguero}, {Samaras}, {Sanchez Gimenez}, {Sanna}, {Santove{\~n}a}, {Sarasso}, {Schultheis}, {Sciacca}, {Segol}, {Segovia}, {S{\'e}gransan}, {Semeux}, {Shahaf}, {Siddiqui}, {Siebert}, {Siltala}, {Slezak}, {Smart}, {Solano}, {Solitro}, {Souami}, {Souchay}, {Spagna}, {Spoto}, {Steele}, {Steidelm{\"u}ller}, {Stephenson}, {S{\"u}veges}, {Szabados}, {Szegedi-Elek}, {Taris}, {Tauran}, {Taylor}, {Teixeira}, {Thuillot}, {Tonello}, {Torra}, {Torra}, {Turon}, {Unger}, {Vaillant}, {van Dillen}, {Vanel}, {Vecchiato}, {Viala}, {Vicente},
  {Voutsinas}, {Weiler}, {Wevers}, {Wyrzykowski}, {Yoldas}, {Yvard}, {Zhao}, {Zorec}, {Zucker}, {Zurbach}, \& {Zwitter}}]{GaiaEDR3-2021a}
{Gaia Collaboration}, {Brown}, A.~G.~A., {Vallenari}, A., {et~al.} 2021, \aap, 649, A1, \dodoi{10.1051/0004-6361/202039657}

\bibitem[{{Gallet} \& {Bouvier}(2015)}]{Gallet2015}
{Gallet}, F., \& {Bouvier}, J. 2015, \aap, 577, A98, \dodoi{10.1051/0004-6361/201525660}

\bibitem[{{Gangi} {et~al.}(2022){Gangi}, {Antoniucci}, {Biazzo}, {Frasca}, {Nisini}, {Alcal{\'a}}, {Giannini}, {Manara}, {Giunta}, {Harutyunyan}, {Munari}, \& {Vitali}}]{Gangi2022}
{Gangi}, M., {Antoniucci}, S., {Biazzo}, K., {et~al.} 2022, \aap, 667, A124, \dodoi{10.1051/0004-6361/202244042}

\bibitem[{{Gillen} {et~al.}(2020){Gillen}, {Hillenbrand}, {Stauffer}, {Aigrain}, {Rebull}, \& {Cody}}]{Gillen2020}
{Gillen}, E., {Hillenbrand}, L.~A., {Stauffer}, J., {et~al.} 2020, \mnras, 495, 1531, \dodoi{10.1093/mnras/staa1016}

\bibitem[{{Gray}(1992)}]{Gray1992}
{Gray}, D.~F. 1992, {The observation and analysis of stellar photospheres.}, Vol.~20 ({Cambridge University Press})

\bibitem[{{Hamilton} {et~al.}(2007){Hamilton}, {Rhode}, \& {Picard}}]{Hamilton2007}
{Hamilton}, C.~M., {Rhode}, K.~L., \& {Picard}, T.~M. 2007, in American Astronomical Society Meeting Abstracts, Vol. 210, American Astronomical Society Meeting Abstracts \#210, 87.08

\bibitem[{{Hartmann}(2003)}]{Hartmann2003}
{Hartmann}, L. 2003, \apj, 585, 398, \dodoi{10.1086/345933}

\bibitem[{{Hartmann} {et~al.}(2016){Hartmann}, {Herczeg}, \& {Calvet}}]{Hartmann2016}
{Hartmann}, L., {Herczeg}, G., \& {Calvet}, N. 2016, \araa, 54, 135, \dodoi{10.1146/annurev-astro-081915-023347}

\bibitem[{{Hartmann} {et~al.}(1986){Hartmann}, {Hewett}, {Stahler}, \& {Mathieu}}]{Hartmann1986}
{Hartmann}, L., {Hewett}, R., {Stahler}, S., \& {Mathieu}, R.~D. 1986, \apj, 309, 275, \dodoi{10.1086/164599}

\bibitem[{{Hayashi}(1961)}]{Hayashi1961}
{Hayashi}, C. 1961, \pasj, 13, 450

\bibitem[{{Healy} {et~al.}(2023){Healy}, {McCullough}, {Schlaufman}, \& {Kovacs}}]{Healy2023}
{Healy}, B.~F., {McCullough}, P.~R., {Schlaufman}, K.~C., \& {Kovacs}, G. 2023, \apj, 944, 39, \dodoi{10.3847/1538-4357/acad7b}

\bibitem[{{Herbst} {et~al.}(2001){Herbst}, {Bailer-Jones}, \& {Mundt}}]{Herbst2001}
{Herbst}, W., {Bailer-Jones}, C.~A.~L., \& {Mundt}, R. 2001, \apjl, 554, L197, \dodoi{10.1086/321706}

\bibitem[{{Herbst} {et~al.}(2002){Herbst}, {Bailer-Jones}, {Mundt}, {Meisenheimer}, \& {Wackermann}}]{Herbst2002}
{Herbst}, W., {Bailer-Jones}, C.~A.~L., {Mundt}, R., {Meisenheimer}, K., \& {Wackermann}, R. 2002, \aap, 396, 513, \dodoi{10.1051/0004-6361:20021362}

\bibitem[{{Herbst} {et~al.}(1994){Herbst}, {Herbst}, {Grossman}, \& {Weinstein}}]{Herbst1994}
{Herbst}, W., {Herbst}, D.~K., {Grossman}, E.~J., \& {Weinstein}, D. 1994, \aj, 108, 1906, \dodoi{10.1086/117204}

\bibitem[{{Herbst} {et~al.}(2000){Herbst}, {Rhode}, {Hillenbrand}, \& {Curran}}]{Herbst2000}
{Herbst}, W., {Rhode}, K.~L., {Hillenbrand}, L.~A., \& {Curran}, G. 2000, \aj, 119, 261, \dodoi{10.1086/301175}

\bibitem[{{Hern{\'a}ndez} {et~al.}(2007){Hern{\'a}ndez}, {Hartmann}, {Megeath}, {Gutermuth}, {Muzerolle}, {Calvet}, {Vivas}, {Brice{\~n}o}, {Allen}, {Stauffer}, {Young}, \& {Fazio}}]{Hernandez2007}
{Hern{\'a}ndez}, J., {Hartmann}, L., {Megeath}, T., {et~al.} 2007, \apj, 662, 1067, \dodoi{10.1086/513735}

\bibitem[{{Hern{\'a}ndez} {et~al.}(2023){Hern{\'a}ndez}, {Zamudio}, {Brice{\~n}o}, {Calvet}, {Zhu}, {Yuan}, {Liu}, {Manzo-Mart{\'\i}nez}, {Rom{\'a}n-Z{\'u}{\~n}iga}, {Serna}, {Mauc{\'o}}, \& {Adame}}]{Hernandez2023}
{Hern{\'a}ndez}, J., {Zamudio}, L.~F., {Brice{\~n}o}, C., {et~al.} 2023, \aj, 165, 205, \dodoi{10.3847/1538-3881/acc467}

\bibitem[{{Hillenbrand} {et~al.}(1998){Hillenbrand}, {Strom}, {Calvet}, {Merrill}, {Gatley}, {Makidon}, {Meyer}, \& {Skrutskie}}]{Hillenbrand1998b}
{Hillenbrand}, L.~A., {Strom}, S.~E., {Calvet}, N., {et~al.} 1998, \aj, 116, 1816, \dodoi{10.1086/300536}

\bibitem[{{Hosokawa} {et~al.}(2011){Hosokawa}, {Offner}, \& {Krumholz}}]{Hosokawa2011}
{Hosokawa}, T., {Offner}, S. S.~R., \& {Krumholz}, M.~R. 2011, \apj, 738, 140, \dodoi{10.1088/0004-637X/738/2/140}

\bibitem[{{Jackson} {et~al.}(2018){Jackson}, {Deliyannis}, \& {Jeffries}}]{Jackson2018}
{Jackson}, R.~J., {Deliyannis}, C.~P., \& {Jeffries}, R.~D. 2018, \mnras, 476, 3245, \dodoi{10.1093/mnras/sty374}

\bibitem[{{Jackson} \& {Jeffries}(2010)}]{Jackson2010a}
{Jackson}, R.~J., \& {Jeffries}, R.~D. 2010, \mnras, 402, 1380, \dodoi{10.1111/j.1365-2966.2009.15983.x}

\bibitem[{{Jackson} {et~al.}(2019){Jackson}, {Jeffries}, {Deliyannis}, {Sun}, \& {Douglas}}]{Jackson2019}
{Jackson}, R.~J., {Jeffries}, R.~D., {Deliyannis}, C.~P., {Sun}, Q., \& {Douglas}, S.~T. 2019, \mnras, 483, 1125, \dodoi{10.1093/mnras/sty3184}

\bibitem[{{Jackson} {et~al.}(2016){Jackson}, {Jeffries}, {Randich}, {Bragaglia}, {Carraro}, {Costado}, {Flaccomio}, {Lanzafame}, {Lardo}, {Monaco}, {Morbidelli}, {Smiljanic}, \& {Zaggia}}]{Jackson2016}
{Jackson}, R.~J., {Jeffries}, R.~D., {Randich}, S., {et~al.} 2016, \aap, 586, A52, \dodoi{10.1051/0004-6361/201527507}

\bibitem[{{Jester} {et~al.}(2005){Jester}, {Schneider}, {Richards}, {Green}, {Schmidt}, {Hall}, {Strauss}, {Vanden Berk}, {Stoughton}, {Gunn}, {Brinkmann}, {Kent}, {Smith}, {Tucker}, \& {Yanny}}]{Jester2005}
{Jester}, S., {Schneider}, D.~P., {Richards}, G.~T., {et~al.} 2005, \aj, 130, 873, \dodoi{10.1086/432466}

\bibitem[{{Jones} \& {Walker}(1988)}]{Jones1988}
{Jones}, B.~F., \& {Walker}, M.~F. 1988, \aj, 95, 1755, \dodoi{10.1086/114773}

\bibitem[{{Kamezaki} {et~al.}(2013){Kamezaki}, {Imura}, {Nagayama}, {Omodaka}, {Handa}, {Yamaguchi}, {Chibueze}, {Sunada}, \& {Nakano}}]{Kamezaki2013}
{Kamezaki}, T., {Imura}, K., {Nagayama}, T., {et~al.} 2013, in Molecular Gas, Dust, and Star Formation in Galaxies, ed. T.~{Wong} \& J.~{Ott}, Vol. 292, 45--45, \dodoi{10.1017/S1743921313000264}

\bibitem[{{K{\"o}nigl}(1991)}]{Konigl1991}
{K{\"o}nigl}, A. 1991, \apjl, 370, L39, \dodoi{10.1086/185972}

\bibitem[{{Kovacs}(2018)}]{Kovacs2018}
{Kovacs}, G. 2018, \aap, 612, L2, \dodoi{10.1051/0004-6361/201731355}

\bibitem[{{Kraus} {et~al.}(2015){Kraus}, {Cody}, {Covey}, {Rizzuto}, {Mann}, \& {Ireland}}]{Kraus2015}
{Kraus}, A.~L., {Cody}, A.~M., {Covey}, K.~R., {et~al.} 2015, \apj, 807, 3, \dodoi{10.1088/0004-637X/807/1/3}

\bibitem[{{Lada}(1985)}]{Lada1985}
{Lada}, C.~J. 1985, \araa, 23, 267, \dodoi{10.1146/annurev.aa.23.090185.001411}

\bibitem[{{Lamm} {et~al.}(2004){Lamm}, {Bailer-Jones}, {Mundt}, {Herbst}, \& {Scholz}}]{Lamm2004}
{Lamm}, M.~H., {Bailer-Jones}, C.~A.~L., {Mundt}, R., {Herbst}, W., \& {Scholz}, A. 2004, \aap, 417, 557, \dodoi{10.1051/0004-6361:20035588}

\bibitem[{{Lamm} {et~al.}(2005){Lamm}, {Mundt}, {Bailer-Jones}, \& {Herbst}}]{Lamm2005}
{Lamm}, M.~H., {Mundt}, R., {Bailer-Jones}, C.~A.~L., \& {Herbst}, W. 2005, \aap, 430, 1005, \dodoi{10.1051/0004-6361:20040492}

\bibitem[{{Lanzafame} {et~al.}(2017){Lanzafame}, {Spada}, \& {Distefano}}]{Lanzafame2017}
{Lanzafame}, A.~C., {Spada}, F., \& {Distefano}, E. 2017, \aap, 597, A63, \dodoi{10.1051/0004-6361/201628833}

\bibitem[{{Levato}(1974)}]{Levato1974}
{Levato}, H. 1974, \aap, 35, 259

\bibitem[{{Lim} {et~al.}(2016){Lim}, {Sung}, {Kim}, {Bessell}, {Hwang}, \& {Park}}]{Lim2016}
{Lim}, B., {Sung}, H., {Kim}, J.~S., {et~al.} 2016, \apj, 831, 116, \dodoi{10.3847/0004-637X/831/2/116}

\bibitem[{{Littlefair} {et~al.}(2010){Littlefair}, {Naylor}, {Mayne}, {Saunders}, \& {Jeffries}}]{Littlefair2010}
{Littlefair}, S.~P., {Naylor}, T., {Mayne}, N.~J., {Saunders}, E.~S., \& {Jeffries}, R.~D. 2010, \mnras, 403, 545, \dodoi{10.1111/j.1365-2966.2010.16066.x}

\bibitem[{{L{\'o}pez-Morales}(2007)}]{Lopez-Morales2007}
{L{\'o}pez-Morales}, M. 2007, \apj, 660, 732, \dodoi{10.1086/513142}

\bibitem[{{Majewski} {et~al.}(2017){Majewski}, {Schiavon}, {Frinchaboy}, {Allende Prieto}, {Barkhouser}, {Bizyaev}, {Blank}, {Brunner}, {Burton}, {Carrera}, {Chojnowski}, {Cunha}, {Epstein}, {Fitzgerald}, {Garc{\'\i}a P{\'e}rez}, {Hearty}, {Henderson}, {Holtzman}, {Johnson}, {Lam}, {Lawler}, {Maseman}, {M{\'e}sz{\'a}ros}, {Nelson}, {Nguyen}, {Nidever}, {Pinsonneault}, {Shetrone}, {Smee}, {Smith}, {Stolberg}, {Skrutskie}, {Walker}, {Wilson}, {Zasowski}, {Anders}, {Basu}, {Beland}, {Blanton}, {Bovy}, {Brownstein}, {Carlberg}, {Chaplin}, {Chiappini}, {Eisenstein}, {Elsworth}, {Feuillet}, {Fleming}, {Galbraith-Frew}, {Garc{\'\i}a}, {Garc{\'\i}a-Hern{\'a}ndez}, {Gillespie}, {Girardi}, {Gunn}, {Hasselquist}, {Hayden}, {Hekker}, {Ivans}, {Kinemuchi}, {Klaene}, {Mahadevan}, {Mathur}, {Mosser}, {Muna}, {Munn}, {Nichol}, {O'Connell}, {Parejko}, {Robin}, {Rocha-Pinto}, {Schultheis}, {Serenelli}, {Shane}, {Silva Aguirre}, {Sobeck}, {Thompson}, {Troup}, {Weinberg}, \& {Zamora}}]{APOGEE2017}
{Majewski}, S.~R., {Schiavon}, R.~P., {Frinchaboy}, P.~M., {et~al.} 2017, \aj, 154, 94, \dodoi{10.3847/1538-3881/aa784d}

\bibitem[{{Makidon} {et~al.}(2004){Makidon}, {Rebull}, {Strom}, {Adams}, \& {Patten}}]{Makidon2004}
{Makidon}, R.~B., {Rebull}, L.~M., {Strom}, S.~E., {Adams}, M.~T., \& {Patten}, B.~M. 2004, \aj, 127, 2228, \dodoi{10.1086/382237}

\bibitem[{{Matt} \& {Pudritz}(2005)}]{Matt2005b}
{Matt}, S., \& {Pudritz}, R.~E. 2005, \apjl, 632, L135, \dodoi{10.1086/498066}

\bibitem[{{Mermilliod} {et~al.}(2009){Mermilliod}, {Mayor}, \& {Udry}}]{Mermilliod2009}
{Mermilliod}, J.~C., {Mayor}, M., \& {Udry}, S. 2009, \aap, 498, 949, \dodoi{10.1051/0004-6361/200810244}

\bibitem[{{Mosser} {et~al.}(2018){Mosser}, {Gehan}, {Belkacem}, {Samadi}, {Michel}, \& {Goupil}}]{Mosser2018}
{Mosser}, B., {Gehan}, C., {Belkacem}, K., {et~al.} 2018, \aap, 618, A109, \dodoi{10.1051/0004-6361/201832777}

\bibitem[{{Mundt} \& {Fried}(1983)}]{Mundt1983}
{Mundt}, R., \& {Fried}, J.~W. 1983, \apjl, 274, L83, \dodoi{10.1086/184155}

\bibitem[{{Nofi} {et~al.}(2021){Nofi}, {Johns-Krull}, {L{\'o}pez-Valdivia}, {Biddle}, {Carvalho}, {Huber}, {Jaffe}, {Llama}, {Mace}, {Prato}, {Skiff}, {Sokal}, {Sullivan}, \& {Tayar}}]{Nofi2021}
{Nofi}, L.~A., {Johns-Krull}, C.~M., {L{\'o}pez-Valdivia}, R., {et~al.} 2021, \apj, 911, 138, \dodoi{10.3847/1538-4357/abeab3}

\bibitem[{{Nordhagen} {et~al.}(2006){Nordhagen}, {Herbst}, {Rhode}, \& {Williams}}]{Nordhagen2006}
{Nordhagen}, S., {Herbst}, W., {Rhode}, K.~L., \& {Williams}, E.~C. 2006, \aj, 132, 1555, \dodoi{10.1086/506985}

\bibitem[{{Oliveira} {et~al.}(2006){Oliveira}, {Jeffries}, {van Loon}, \& {Rushton}}]{Oliveira2006}
{Oliveira}, J.~M., {Jeffries}, R.~D., {van Loon}, J.~T., \& {Rushton}, M.~T. 2006, \mnras, 369, 272, \dodoi{10.1111/j.1365-2966.2006.10299.x}

\bibitem[{{Ostriker} \& {Shu}(1995)}]{Ostriker1995}
{Ostriker}, E.~C., \& {Shu}, F.~H. 1995, \apj, 447, 813, \dodoi{10.1086/175920}

\bibitem[{{Palla} \& {Stahler}(2000)}]{Palla2000}
{Palla}, F., \& {Stahler}, S.~W. 2000, \apj, 540, 255, \dodoi{10.1086/309312}

\bibitem[{{Park} {et~al.}(2000){Park}, {Sung}, {Bessell}, \& {Kang}}]{Park2000}
{Park}, B.-G., {Sung}, H., {Bessell}, M.~S., \& {Kang}, Y.~H. 2000, \aj, 120, 894, \dodoi{10.1086/301459}

\bibitem[{{Pecaut} \& {Mamajek}(2013)}]{Pecaut2013}
{Pecaut}, M.~J., \& {Mamajek}, E.~E. 2013, \apjs, 208, 9, \dodoi{10.1088/0067-0049/208/1/9}

\bibitem[{{Pecaut} {et~al.}(2012){Pecaut}, {Mamajek}, \& {Bubar}}]{Pecaut2012}
{Pecaut}, M.~J., {Mamajek}, E.~E., \& {Bubar}, E.~J. 2012, \apj, 746, 154, \dodoi{10.1088/0004-637X/746/2/154}

\bibitem[{{P{\'e}rez Paolino} {et~al.}(2024){P{\'e}rez Paolino}, {Bary}, {Hillenbrand}, \& {Markham}}]{Paolino2024}
{P{\'e}rez Paolino}, F., {Bary}, J.~S., {Hillenbrand}, L.~A., \& {Markham}, M. 2024, \apj, 967, 45, \dodoi{10.3847/1538-4357/ad393b}

\bibitem[{{Pfalzner} {et~al.}(2022){Pfalzner}, {Dehghani}, \& {Michel}}]{Pfalzner2022}
{Pfalzner}, S., {Dehghani}, S., \& {Michel}, A. 2022, \apjl, 939, L10, \dodoi{10.3847/2041-8213/ac9839}

\bibitem[{{Pfalzner} \& {Dincer}(2024)}]{Pfalzner2024}
{Pfalzner}, S., \& {Dincer}, F. 2024, \apj, 963, 122, \dodoi{10.3847/1538-4357/ad1bef}

\bibitem[{{Pinsonneault} {et~al.}(1990){Pinsonneault}, {Kawaler}, \& {Demarque}}]{Pinsonneault1990}
{Pinsonneault}, M.~H., {Kawaler}, S.~D., \& {Demarque}, P. 1990, \apjs, 74, 501, \dodoi{10.1086/191507}

\bibitem[{{Pinsonneault} {et~al.}(1989){Pinsonneault}, {Kawaler}, {Sofia}, \& {Demarque}}]{Pinsonneault1989}
{Pinsonneault}, M.~H., {Kawaler}, S.~D., {Sofia}, S., \& {Demarque}, P. 1989, \apj, 338, 424, \dodoi{10.1086/167210}

\bibitem[{{Ram{\'\i}rez} {et~al.}(2004){Ram{\'\i}rez}, {Rebull}, {Stauffer}, {Hearty}, {Hillenbrand}, {Jones}, {Makidon}, {Pravdo}, {Strom}, \& {Werner}}]{Ramirez2004}
{Ram{\'\i}rez}, S.~V., {Rebull}, L., {Stauffer}, J., {et~al.} 2004, \aj, 127, 2659, \dodoi{10.1086/383290}

\bibitem[{{Rebull}(2001)}]{Rebull2001}
{Rebull}, L.~M. 2001, \aj, 121, 1676, \dodoi{10.1086/319393}

\bibitem[{{Rebull} {et~al.}(2020){Rebull}, {Stauffer}, {Cody}, {Hillenbrand}, {Bouvier}, {Roggero}, \& {David}}]{Rebull2020}
{Rebull}, L.~M., {Stauffer}, J.~R., {Cody}, A.~M., {et~al.} 2020, \aj, 159, 273, \dodoi{10.3847/1538-3881/ab893c}

\bibitem[{{Rebull} {et~al.}(2018){Rebull}, {Stauffer}, {Cody}, {Hillenbrand}, {David}, \& {Pinsonneault}}]{Rebull2018}
---. 2018, \aj, 155, 196, \dodoi{10.3847/1538-3881/aab605}

\bibitem[{{Rebull} {et~al.}(2004){Rebull}, {Wolff}, \& {Strom}}]{Rebull2004}
{Rebull}, L.~M., {Wolff}, S.~C., \& {Strom}, S.~E. 2004, \aj, 127, 1029, \dodoi{10.1086/380931}

\bibitem[{{Rebull} {et~al.}(2002){Rebull}, {Makidon}, {Strom}, {Hillenbrand}, {Birmingham}, {Patten}, {Jones}, {Yagi}, \& {Adams}}]{Rebull2002a}
{Rebull}, L.~M., {Makidon}, R.~B., {Strom}, S.~E., {et~al.} 2002, \aj, 123, 1528, \dodoi{10.1086/338904}

\bibitem[{{Rhode} {et~al.}(2001){Rhode}, {Herbst}, \& {Mathieu}}]{Rhode2001b}
{Rhode}, K.~L., {Herbst}, W., \& {Mathieu}, R.~D. 2001, \aj, 122, 3258, \dodoi{10.1086/324448}

\bibitem[{{Riello} {et~al.}(2021){Riello}, {De Angeli}, {Evans}, {Montegriffo}, {Carrasco}, {Busso}, {Palaversa}, {Burgess}, {Diener}, {Davidson}, {Rowell}, {Fabricius}, {Jordi}, {Bellazzini}, {Pancino}, {Harrison}, {Cacciari}, {van Leeuwen}, {Hambly}, {Hodgkin}, {Osborne}, {Altavilla}, {Barstow}, {Brown}, {Castellani}, {Cowell}, {De Luise}, {Gilmore}, {Giuffrida}, {Hidalgo}, {Holland}, {Marinoni}, {Pagani}, {Piersimoni}, {Pulone}, {Ragaini}, {Rainer}, {Richards}, {Sanna}, {Walton}, {Weiler}, \& {Yoldas}}]{GaiaEDR3Conv2021}
{Riello}, M., {De Angeli}, F., {Evans}, D.~W., {et~al.} 2021, \aap, 649, A3, \dodoi{10.1051/0004-6361/202039587}

\bibitem[{{Serna} {et~al.}(2021){Serna}, {Hernandez}, {Kounkel}, {Manzo-Mart{\'\i}nez}, {Roman-Lopes}, {Rom{\'a}n-Z{\'u}{\~n}iga}, {Gracia Batista}, {Pinz{\'o}n}, {Calvet}, {Brice{\~n}o}, {Tapia}, {Su{\'a}rez}, {Pe{\~n}a Ram{\'\i}rez}, {Stassun}, {Covey}, {Vargas-Gonz{\'a}lez}, \& {Fern{\'a}ndez-Trincado}}]{Serna2021}
{Serna}, J., {Hernandez}, J., {Kounkel}, M., {et~al.} 2021, \apj, 923, 177, \dodoi{10.3847/1538-4357/ac300a}

\bibitem[{{Serna} {et~al.}(2024){Serna}, {Pinz{\'o}n}, {Hern{\'a}ndez}, {Manzo-Mart{\'\i}nez}, {Mauco}, {Rom{\'a}n-Z{\'u}{\~n}iga}, {Calvet}, {Brice{\~n}o}, {L{\'o}pez-Valdivia}, {Kounkel}, {Stringfellow}, {Stassun}, {Pinsonneault}, {Adame}, {Cao}, {Covey}, {Bayo}, {Roman-Lopes}, {Nitschelm}, \& {Lane}}]{Serna2024}
{Serna}, J., {Pinz{\'o}n}, G., {Hern{\'a}ndez}, J., {et~al.} 2024, \apj, 968, 68, \dodoi{10.3847/1538-4357/ad3a6b}

\bibitem[{{Shu} {et~al.}(1994){Shu}, {Najita}, {Ostriker}, {Wilkin}, {Ruden}, \& {Lizano}}]{Shu1994a}
{Shu}, F., {Najita}, J., {Ostriker}, E., {et~al.} 1994, \apj, 429, 781, \dodoi{10.1086/174363}

\bibitem[{{Smith} {et~al.}(2021){Smith}, {Gillen}, {Queloz}, {Hillenbrand}, {Acton}, {Alves}, {Anderson}, {Bayliss}, {Briegal}, {Burleigh}, {Casewell}, {Delrez}, {Dransfield}, {Ducrot}, {Gill}, {Gillon}, {Goad}, {G{\"u}nther}, {Henderson}, {Jenkins}, {Jehin}, {Moyano}, {Murray}, {Pedersen}, {Sebastian}, {Thompson}, {Tilbrook}, {Triaud}, {Vines}, \& {Wheatley}}]{Smith2021}
{Smith}, G.~D., {Gillen}, E., {Queloz}, D., {et~al.} 2021, \mnras, 507, 5991, \dodoi{10.1093/mnras/stab2374}

\bibitem[{{Somers} {et~al.}(2020){Somers}, {Cao}, \& {Pinsonneault}}]{Somers2020}
{Somers}, G., {Cao}, L., \& {Pinsonneault}, M.~H. 2020, \apj, 891, 29, \dodoi{10.3847/1538-4357/ab722e}

\bibitem[{{Somers} \& {Pinsonneault}(2015)}]{Somers2015a}
{Somers}, G., \& {Pinsonneault}, M.~H. 2015, \mnras, 449, 4131, \dodoi{10.1093/mnras/stv630}

\bibitem[{{Stassun} {et~al.}(1999){Stassun}, {Mathieu}, {Mazeh}, \& {Vrba}}]{Stassun1999}
{Stassun}, K.~G., {Mathieu}, R.~D., {Mazeh}, T., \& {Vrba}, F.~J. 1999, \aj, 117, 2941, \dodoi{10.1086/300881}

\bibitem[{{Stauffer} {et~al.}(2018){Stauffer}, {Rebull}, {Cody}, {Hillenbrand}, {Pinsonneault}, {Barrado}, {Bouvier}, \& {David}}]{Stauffer2018}
{Stauffer}, J., {Rebull}, L.~M., {Cody}, A.~M., {et~al.} 2018, \aj, 156, 275, \dodoi{10.3847/1538-3881/aae9ec}

\bibitem[{{Stauffer} {et~al.}(2003){Stauffer}, {Jones}, {Backman}, {Hartmann}, {Barrado y Navascu{\'e}s}, {Pinsonneault}, {Terndrup}, \& {Muench}}]{Stauffer2003}
{Stauffer}, J.~R., {Jones}, B.~F., {Backman}, D., {et~al.} 2003, \aj, 126, 833, \dodoi{10.1086/376739}

\bibitem[{{Sung} {et~al.}(1997){Sung}, {Bessell}, \& {Lee}}]{Sung1997}
{Sung}, H., {Bessell}, M.~S., \& {Lee}, S.-W. 1997, \aj, 114, 2644, \dodoi{10.1086/118674}

\bibitem[{{Tody}(1986)}]{Tody1986}
{Tody}, D. 1986, in Society of Photo-Optical Instrumentation Engineers (SPIE) Conference Series, Vol. 627, Instrumentation in astronomy VI, ed. D.~L. {Crawford}, 733, \dodoi{10.1117/12.968154}

\bibitem[{{Tody}(1993)}]{Tody1993}
{Tody}, D. 1993, in Astronomical Society of the Pacific Conference Series, Vol.~52, Astronomical Data Analysis Software and Systems II, ed. R.~J. {Hanisch}, R.~J.~V. {Brissenden}, \& J.~{Barnes}, 173

\bibitem[{{Tonry} \& {Davis}(1979)}]{Tonry1979}
{Tonry}, J., \& {Davis}, M. 1979, \aj, 84, 1511, \dodoi{10.1086/112569}

\bibitem[{{Torres} {et~al.}(2010){Torres}, {Andersen}, \& {Gim{\'e}nez}}]{Torres2010}
{Torres}, G., {Andersen}, J., \& {Gim{\'e}nez}, A. 2010, \aapr, 18, 67, \dodoi{10.1007/s00159-009-0025-1}

\bibitem[{{Turner}(2012)}]{Turner2012}
{Turner}, D.~G. 2012, Astronomische Nachrichten, 333, 174, \dodoi{10.1002/asna.201111643}

\bibitem[{{Venuti} {et~al.}(2019){Venuti}, {Damiani}, \& {Prisinzano}}]{Venuti2019}
{Venuti}, L., {Damiani}, F., \& {Prisinzano}, L. 2019, \aap, 621, A14, \dodoi{10.1051/0004-6361/201833253}

\bibitem[{{Venuti} {et~al.}(2014){Venuti}, {Bouvier}, {Flaccomio}, {Alencar}, {Irwin}, {Stauffer}, {Cody}, {Teixeira}, {Sousa}, {Micela}, {Cuillandre}, \& {Peres}}]{Venuti2014}
{Venuti}, L., {Bouvier}, J., {Flaccomio}, E., {et~al.} 2014, \aap, 570, A82, \dodoi{10.1051/0004-6361/201423776}

\bibitem[{{Venuti} {et~al.}(2017){Venuti}, {Bouvier}, {Cody}, {Stauffer}, {Micela}, {Rebull}, {Alencar}, {Sousa}, {Hillenbrand}, \& {Flaccomio}}]{Venuti2017}
{Venuti}, L., {Bouvier}, J., {Cody}, A.~M., {et~al.} 2017, \aap, 599, A23, \dodoi{10.1051/0004-6361/201629537}

\bibitem[{{Venuti} {et~al.}(2018){Venuti}, {Prisinzano}, {Sacco}, {Flaccomio}, {Bonito}, {Damiani}, {Micela}, {Guarcello}, {Randich}, {Stauffer}, {Cody}, {Jeffries}, {Alencar}, {Alfaro}, {Lanzafame}, {Pancino}, {Bayo}, {Carraro}, {Costado}, {Frasca}, {Jofr{\'e}}, {Morbidelli}, {Sousa}, \& {Zaggia}}]{Venuti2018}
{Venuti}, L., {Prisinzano}, L., {Sacco}, G.~G., {et~al.} 2018, \aap, 609, A10, \dodoi{10.1051/0004-6361/201731103}

\bibitem[{{Vogel} \& {Kuhi}(1981)}]{Vogel1981}
{Vogel}, S.~N., \& {Kuhi}, L.~V. 1981, \apj, 245, 960, \dodoi{10.1086/158872}

\bibitem[{{Walter}(1987)}]{Walter1987}
{Walter}, F.~M. 1987, \pasp, 99, 31, \dodoi{10.1086/131952}

\bibitem[{{Wenger} {et~al.}(2000){Wenger}, {Ochsenbein}, {Egret}, {Dubois}, {Bonnarel}, {Borde}, {Genova}, {Jasniewicz}, {Lalo{\"e}}, {Lesteven}, \& {Monier}}]{SIMBAD2000}
{Wenger}, M., {Ochsenbein}, F., {Egret}, D., {et~al.} 2000, \aaps, 143, 9, \dodoi{10.1051/aas:2000332}

\bibitem[{{Zahn} \& {Bouchet}(1989)}]{Zahn1989}
{Zahn}, J.~P., \& {Bouchet}, L. 1989, \aap, 223, 112

\end{thebibliography}
\bibliographystyle{aasjournal}

\end{document}